\documentclass[manuscript]{acmart}

\setcopyright{none}

\usepackage{amsmath,amsfonts}
\usepackage{graphicx}
\usepackage{textcomp}
\usepackage{xcolor}
\def\BibTeX{{\rm B\kern-.05em{\sc i\kern-.025em b}\kern-.08em
    T\kern-.1667em\lower.7ex\hbox{E}\kern-.125emX}}

\usepackage[utf8]{inputenc}
\usepackage[T1]{fontenc}
\usepackage{microtype}
\usepackage{subcaption}
\usepackage{multirow}
\usepackage{booktabs}
\usepackage{enumitem}
\usepackage{makecell}
\usepackage{pifont}
\usepackage{xspace}
\usepackage[most]{tcolorbox}

\newcommand{\ie}{{\em i.e.},\xspace}
\newcommand{\eg}{{\em e.g.},\xspace}
\newcommand{\MIG}{{\texttt{SeeR}}\xspace}

\newtcolorbox{promptbox}[1]{
    enhanced,
    colback=gray!4,
    colframe=black!65,
    colbacktitle=gray!14,
    coltitle=black,
    title={#1},
    fonttitle=\small\bfseries,
    fontupper=\footnotesize,
    halign=left,
    boxrule=0.6pt,
    arc=1mm,
    left=1.5mm,
    right=1.5mm,
    top=1.2mm,
    bottom=1.2mm
}

\begin{document}

\title{Spatial Organization Is Accessibility Structure: Hierarchical Screen Reading of Mobile UIs}

\author{Yongxiang Hu}
\affiliation{
  \institution{College of Computer Science and Artificial Intelligence, Fudan University}
  \city{Shanghai}
  \country{China}
}

\author{Xuan Wang}
\affiliation{
  \institution{College of Computer Science and Artificial Intelligence, Fudan University}
  \city{Shanghai}
  \country{China}
}

\author{Qingyang Qiu}
\affiliation{
  \institution{College of Computer Science and Artificial Intelligence, Fudan University}
  \city{Shanghai}
  \country{China}
}
\email{qyqiu25@m.fudan.edu.cn}

\author{Yingjie Liu}
\affiliation{
  \institution{College of Computer Science and Artificial Intelligence, Fudan University}
  \city{Shanghai}
  \country{China}
}

\author{Chun Kang}
\affiliation{
  \institution{College of Computer Science and Artificial Intelligence, Fudan University}
  \city{Shanghai}
  \country{China}
}

\author{Ruiying Zeng}
\affiliation{
  \institution{College of Computer Science and Artificial Intelligence, Fudan University}
  \city{Shanghai}
  \country{China}
}
\email{ryzeng22@m.fudan.edu.cn}

\author{Xin Wang}
\affiliation{
  \institution{College of Computer Science and Artificial Intelligence, Fudan University}
  \city{Shanghai}
  \country{China}
}

\author{Yangfan Zhou}
\affiliation{
  \institution{College of Computer Science and Artificial Intelligence, Fudan University}
  \city{Shanghai}
  \country{China}
}

\begin{teaserfigure}
    \centering
    \begin{subfigure}[t]{0.49\textwidth}
        \centering
        \includegraphics[width=\linewidth]{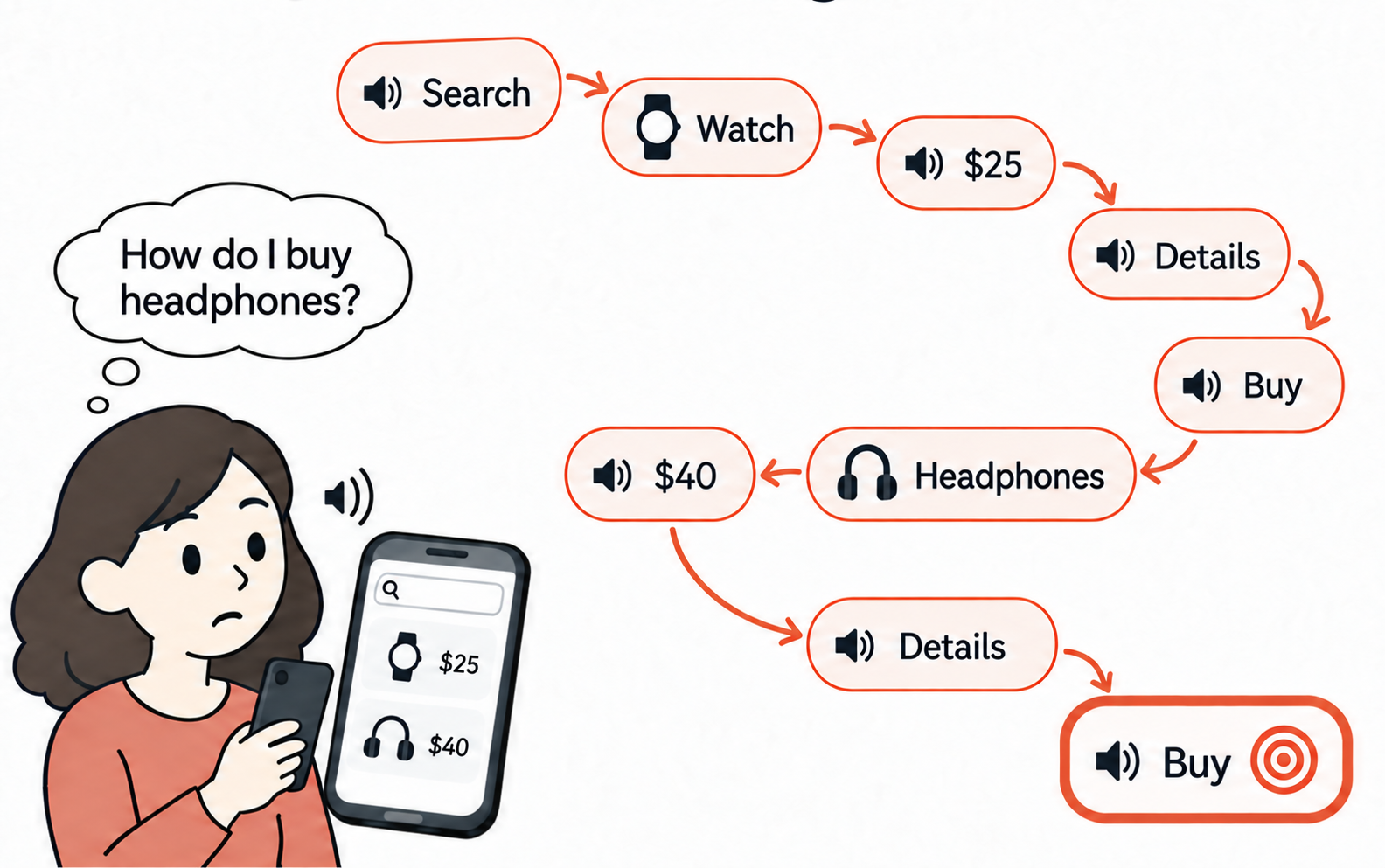}
        \caption{Sequential screen reading}
        \label{fig:teaser-sequential}
    \end{subfigure}\hfill
    \begin{subfigure}[t]{0.49\textwidth}
        \centering
        \includegraphics[width=\linewidth]{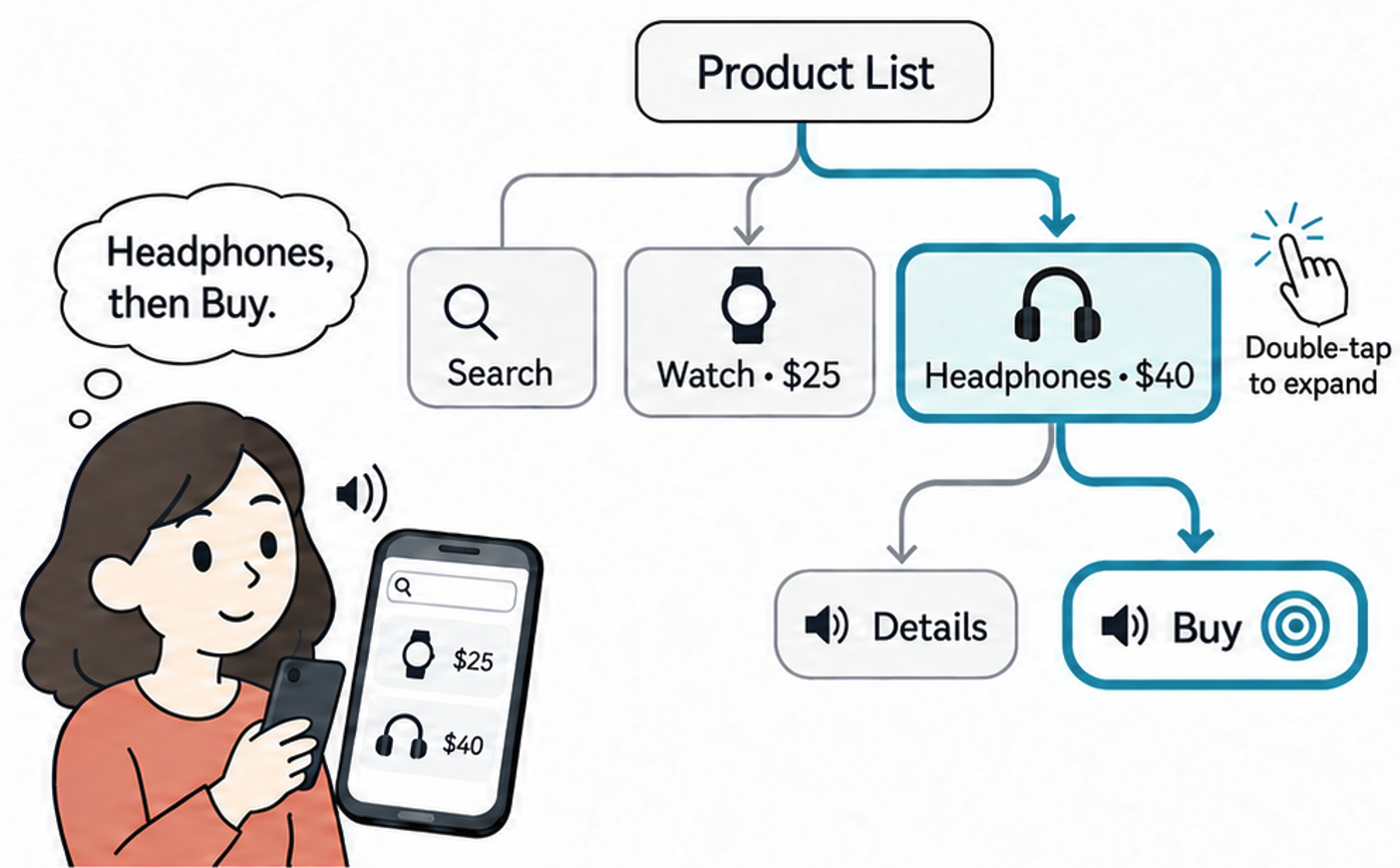}
        \caption{Hierarchical screen reading}
        \label{fig:teaser-hierarchical}
    \end{subfigure}
    \caption{Sequential screen reading and hierarchical screen reading of the same mobile shopping interface. Sequential screen reading requires traversing fragment elements to locate a target control. Hierarchical reading enables users to browse concise functional units and expand a selected unit to access its details and controls.}
    \Description{Two panels compare locating the Buy control for headphones. On the left, a screen-reader user follows a linear sequence through search, watch information and controls, and then headphone information and controls. On the right, a hierarchy groups the same interface into Search, Watch, and Headphones under Product List. The highlighted route follows Headphones and expands it to expose Details and Buy.}
    \label{fig:teaser}
\end{teaserfigure}

\begin{abstract}
Screen readers for blind and low-vision (BLV) users flatten mobile UIs into a linear sequence. Modern mobile UIs, however, contain dozens of elements grouped by spatial layout. Such linear access is ill-suited for BLV users, who must infer each element's meaning and which elements form a functional unit. Drawing on a quantitative TalkBack audit and formative interviews, we present Spatial Screen Reader (\MIG), which builds a function-aware hierarchy from accessibility metadata and provides unit summaries with hierarchical navigation. Compared with TalkBack, \MIG reduces non-informative focus stops by 82.1\% and navigation steps by 52.7\%. With ten BLV participants, summaries helped them connect controls to their functions and skip nonessential elements, especially on unfamiliar pages, though extra navigation levels added effort when the target was already known. These findings recast spatial organization as accessibility information: apps and platforms should expose it, rather than leaving screen readers to infer it post hoc.
\end{abstract}

\ccsdesc[500]{Human-centered computing~Accessibility}
\ccsdesc[500]{Human-centered computing~Interaction design}

\keywords{Screen readers, Accessibility, Mobile interfaces}

\maketitle

\section{Introduction}
\label{sec:intro}

Screen readers make mobile user interfaces (UIs) accessible through speech, focus navigation, and gesture commands~\cite{url:apple-voiceover,url:android-talkback}.
Although effective, screen readers typically provide element-by-element, sequential access. This leads BLV users to understand an interface differently from visual inspection.
Visual UIs organize elements through spatial grouping and hierarchy, allowing sighted users to quickly grasp the overall semantics of a page.
Screen readers, in contrast, largely linearize these elements into a flattened sequence of individual announcements, leaving their relationships implicit. 
As a result, even when all elements are correctly ordered and announced, reconstructing the overall semantics of the UI page remains difficult for
BLV users~\cite{Ramakrishnan2017NonvisualBrowsing}.

\begin{figure*}[t]
    \centering
        \begin{subfigure}[t]{0.32\textwidth}
        \centering
        \includegraphics[width=\linewidth]{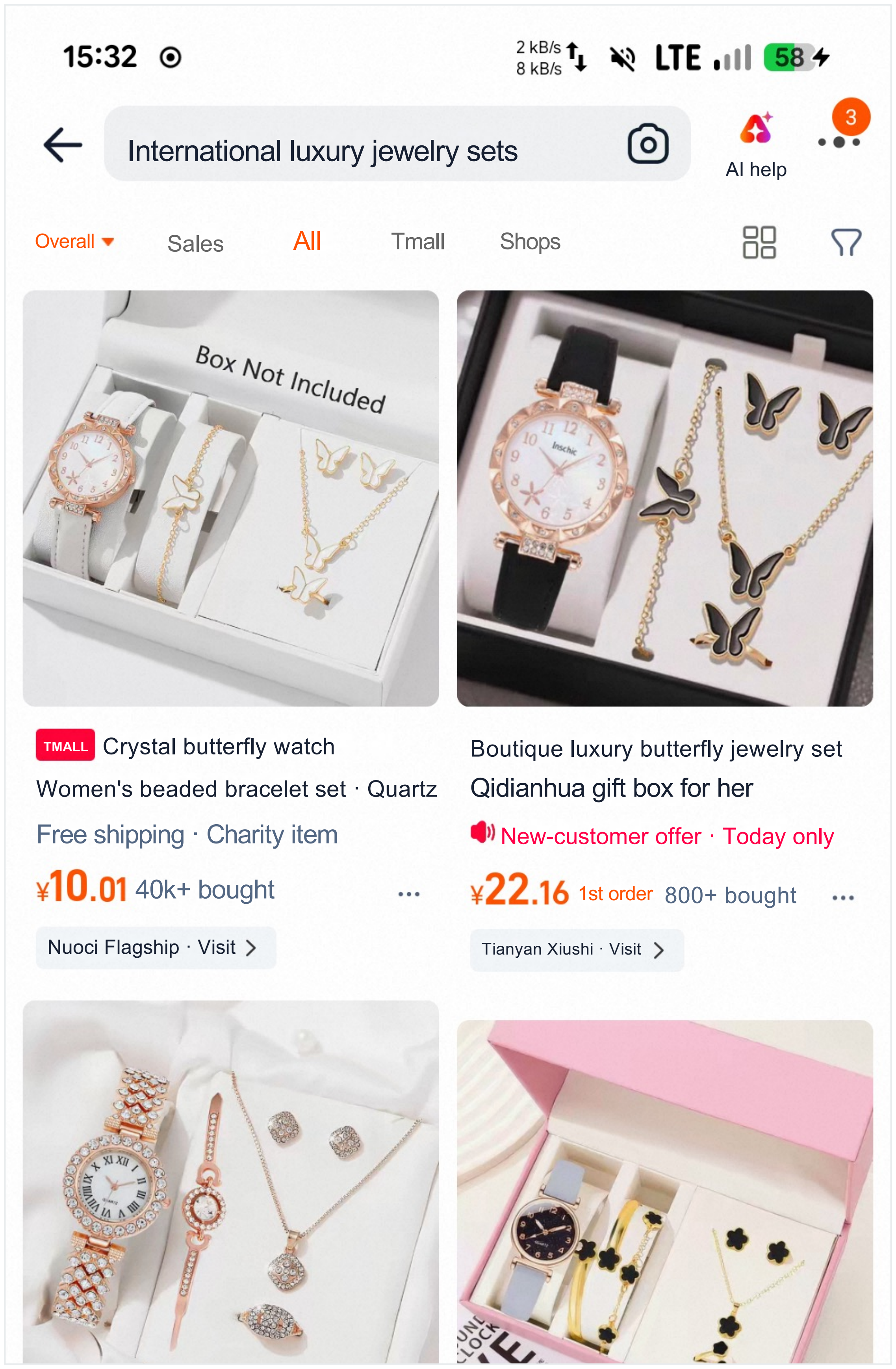}
        \caption{Shopping card}
        \label{ffig:shopping_card}
    \end{subfigure}\hfill
    \begin{subfigure}[t]{0.32\textwidth}
        \centering
        \includegraphics[width=\linewidth]{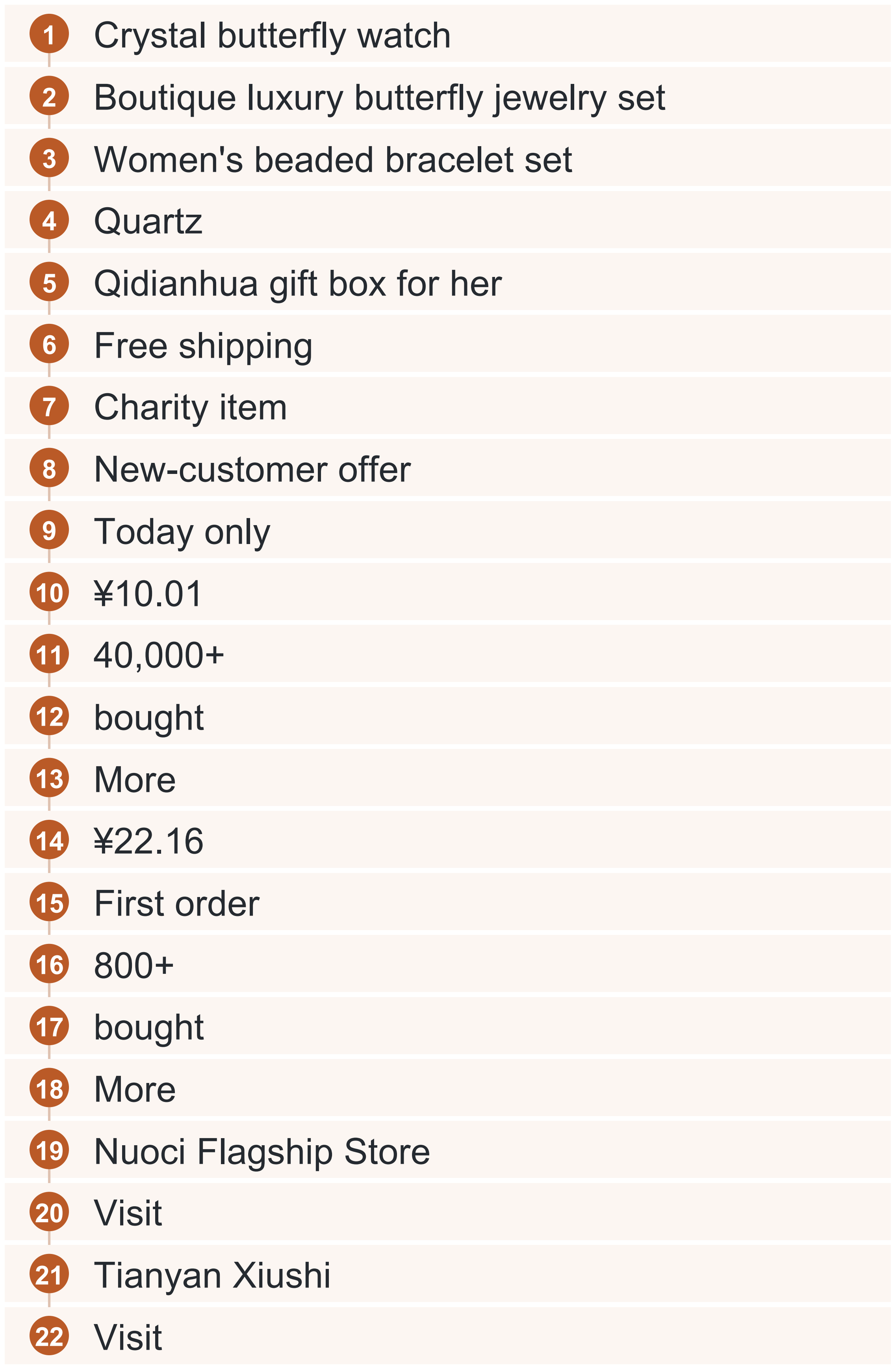}
        \caption{Sequential reading}
        \label{ffig:sequential_reading}
    \end{subfigure}\hfill
    \begin{subfigure}[t]{0.32\textwidth}
        \centering
        \includegraphics[width=\linewidth]{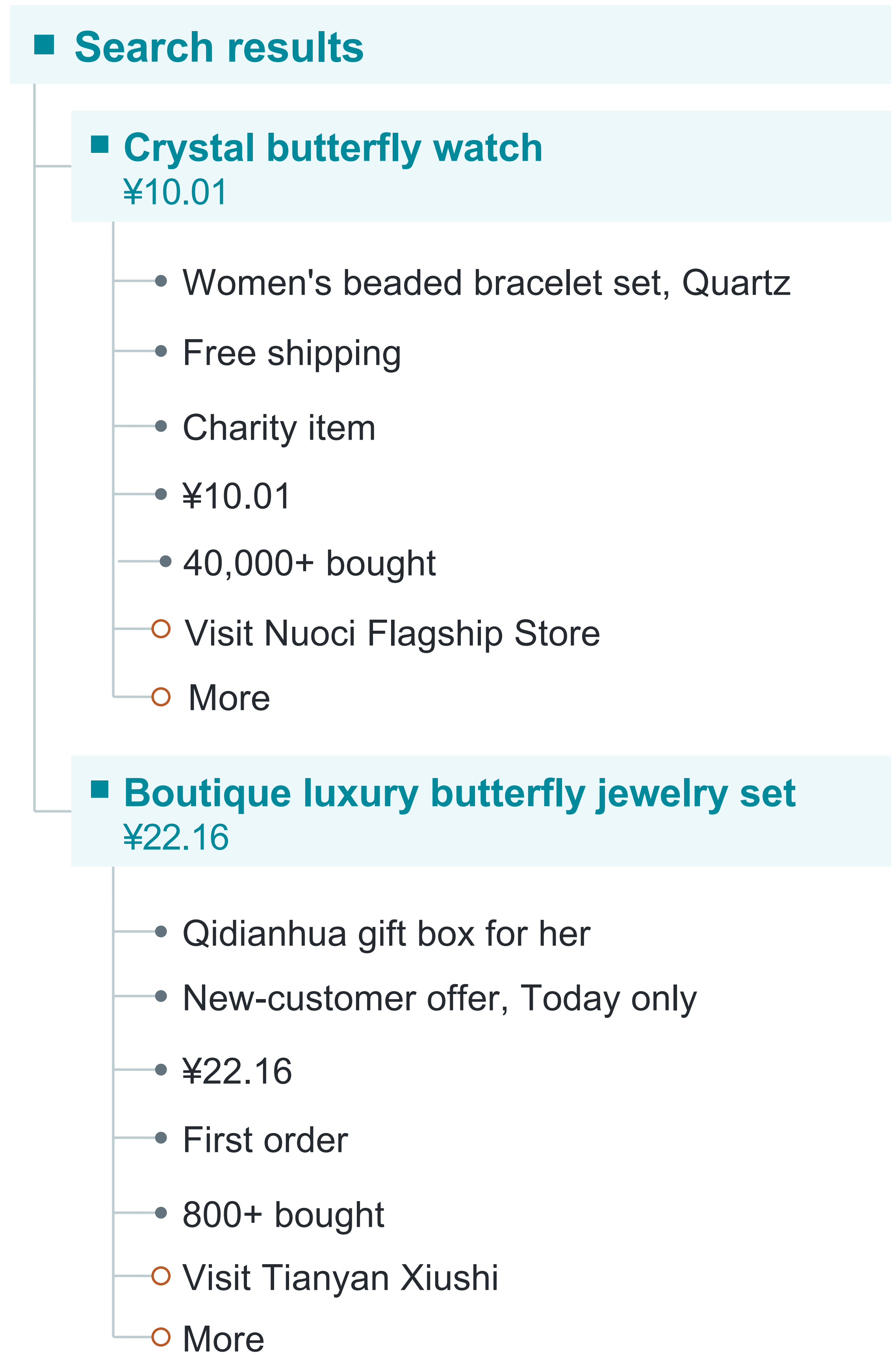}
        \caption{Hierarchical reading}
        \label{ffig:hierarchical_reading}
    \end{subfigure}
    \caption{A shopping card and its reading representations. Sequential reading presents the card's elements one by one. Hierarchical reading presents a summary of the functional unit and provides access to its elements on demand.}
    \label{ffig:1_example}
    \Description{Three panels arranged from left to right show a shopping card screenshot, its sequential reading output, and its hierarchical reading output.}
\end{figure*}

Modern mobile interfaces further amplify this difficulty, as they increasingly organize dense elements into functional units such as cards, lists, and menus~\cite{Deka2017Rico,Liu2018DesignSemantics}.
For example, as shown in Figure~\ref{ffig:shopping_card},a shopping card, typically combines elements such as a product name, image, rating, price, promotion, and action button to represent a single product.
Although spatial structure makes these elements visually coherent as a functional unit, BLV users must infer the same associations solely from the spoken content and reading order, which requires substantially greater cognitive effort~\cite{UbiComp2022SemanticNav}.
Such burden is prevalent in practice. 
Our quantitative study in Section~\ref{ssec:quantitative-audit} shows that 96.0\% of pages contain functional units, with an average of 76.8\% of elements per page belonging to such units.
Moreover, 17 of 50 pages (34.0\%) exhibit functional-unit interruptions during screen-reader navigation, disrupting these structural relationships.

There have been many attempts to improve the quality of screen-reader output, but to our knowledge, none of them targets the fragmentation of UI elements.
Several tools provide hierarchical access to structured content such as charts and diagrams, but they rely on hierarchies already encoded in source data~\cite{chi2023ChartReader,chi2024TADA}.
Other attempts mostly focus on accessibility repair, which recover missing labels, roles, and controls but still present general UIs as separate elements~\cite{Zhang2017InteractionProxies,Zhang2018RobustAnnotation,Zhang2021ScreenRecognition,IMWUT2025TapNav}.
More recently, LLM-based tools have provided descriptions of entire pages or visual content~\cite{chi2026AskEase,chi2026GeoVisA11y,GonzalezPenuela2026VisualAccess}.
However, these descriptions do not reconstruct the relationships among UI elements.
Across these approaches, screen-reader users still lack a structured representation of how individual UI elements work together to form UI functions.

Therefore, this paper investigates whether a function-aware hierarchical representation can support blind and low-vision users' understanding of modern, content-dense mobile UIs.
Specifically, we develop \textit{Spatial Screen Reader} (\MIG), which replaces flat element-by-element reading with hierarchical access organized around functional units.
To construct these hierarchies, \MIG analyzes accessibility metadata and spatial bounds to group UI elements into functional units and organize them top down.
Then, a large language model (LLM) is prompted to produce a concise semantic description for each of these units.
During reading, \MIG initially exposes the top level of the hierarchy, allowing users to move among functional units, expand a unit to inspect its contents, and activate the interactive UI elements.
For example, as shown in Figure~\ref{ffig:1_example}, instead of reading every element in a shopping card sequentially, \MIG announces only the product name and price, omitting other details unless the user requests them with a long press.
This interaction pattern provides access to the elements retained in the hierarchy while reducing the number of elements users must hear before they can understand.

To investigate the effectiveness of \MIG, we conducted a technical evaluation and a user study.
As for the technical evaluation, results on 50 UI pages drawn from 10 apps showed that \MIG achieved a pairwise F1 of 0.77 for functional-unit identification, with a median hierarchy construction latency of 5.8 ms.
These results demonstrate that \MIG can accurately and efficiently transform mobile UIs into hierarchies of functional units.
Moreover, in our user study with ten BLV participants, the hierarchical screen-reader access increased the task-completion rate from 91.9\% to 97.3\%, reduced the mean number of nodes visited by 45.8\% relative to TalkBack, and received an average subjective rating of 3.73/5 across the eight questionnaire items.
We discuss the practical value and limitations of function-aware hierarchical reading and outline directions for improving screen-reader interaction.

We summarize this paper's contributions as follows:
\begin{itemize}
    \item We are the first to pinpoint the loss of spatial relationships among UI elements as a key barrier to understanding mobile UIs through screen readers.
    \item We introduce function-aware hierarchy as the first non-visual representation of mobile UIs that preserves their spatial organization, and implement \MIG\footnote{The source code is available at \url{https://anonymous.4open.science/r/reader-artifact-4115}.}, which automatically constructs such a hierarchy from accessibility metadata and spatial coordinates.
    \item Our technical evaluation and user study show that function-aware hierarchical reading reduces listening effort and supports understanding of mobile UIs. Additionally, user feedback provides valuable insights into improving future screen readers.
\end{itemize}

\section{Related Work}
\label{sec:related}

Screen readers rely on accessible representations to convey UI content through speech~\cite{url:apple-voiceover,url:android-talkback}.
However, empirical studies have repeatedly found missing semantics, inaccessible components, and problematic navigation structures in real-world mobile apps~\cite{Yan2019CurrentStatus,Alshayban2020Accessibility,Chen2022Accessible,Fok2022MissingLabels,Bi2022AccessibilityPractice}.
Prior work has consequently explored element-level accessibility enhancement, structured non-visual navigation, and the recovery of higher-order structure from mobile GUIs~\cite{Zhang2017InteractionProxies,Zhang2021ScreenRecognition,chi2023ChartReader,Billah2017SpeedDial,Wu2021ScreenParsing,Xie2022PerceptualGrouping,Zhang2024DontConfuse}.
We review these three lines of work below.

\subsection{UI Elements Accessibility Enhancement}

Mobile screen readers rely on element-level accessibility information to identify UI components, announce their content and state, and expose available actions~\cite{url:android-accessibility-node,url:apple-voiceover-hig}.
When this information is incomplete, prior work has enhanced different aspects of an element's accessible representation.
A substantial line of research recovers missing semantic content, such as labels, descriptions, and input hints~\cite{Chen2020LabelDroid,Li2020WidgetCaptioning,Mehralian2021COALA,Liu2024HintDroid,Zhong2025Caption}.
Beyond textual descriptions, Screen Recognition infers richer accessibility metadata, including UI content, state, and interactivity, directly from pixels~\cite{Zhang2021ScreenRecognition}, while recent LLM-based approaches further recover visually encoded component states~\cite{Zhang2025ComponentStates}.
Other work targets interaction accessibility: Interaction Proxies repairs otherwise inaccessible interactions at runtime~\cite{Zhang2017InteractionProxies}, and Robust Annotation makes such repairs more resilient to interface changes~\cite{Zhang2018RobustAnnotation}.

Together, these approaches make individual UI elements easier to identify, understand, and operate.
Their enhancements, however, remain primarily at the granularity of individual elements: they improve what each element conveys or enables, but do not determine how multiple elements form a larger functional unit.
Thus, even when individual elements are correctly exposed, users may still need to traverse them one by one.

\subsection{Structured Non-visual Navigation}
Beyond enhancing individual elements, prior work has sought to reduce the cost of sequential screen-reader navigation by changing the granularity of access, allowing users to navigate larger structured units before inspecting individual elements.
For structured visual content such as charts and diagrams, these units and their relationships are often encoded in the underlying data or topology, enabling overview-to-detail and relation-aware navigation~\cite{Sharif2022VoxLens,chi2023ChartReader,chi2024Customization,chi2024TADA,chi2026Arboretum}.
Chart Reader, for example, lets users move between an overview of a chart and its detailed information~\cite{chi2023ChartReader}, while TADA preserves node-link relationships through touch and audio-based interaction~\cite{chi2024TADA}.

Web screen reading has similarly reduced one-by-one traversal by grouping related webpage elements into meaningful regions~\cite{Ashok2017SRAA,Billah2017SpeedDial,Ferdous2021STag}.
SRAA groups related HTML elements into semantic units such as search results, menus, and widgets, allowing users to interact with these units rather than traverse every underlying element~\cite{Ashok2017SRAA}.
Speed-Dial further organizes such units into a hierarchical model, allowing users to first locate a relevant region and then inspect its contents~\cite{Billah2017SpeedDial}.
Related work provides semantic tables of contents or spatially informed browsing to reduce repeated serial navigation~\cite{Ferdous2021STag,ChhedaKothary2025SpatialBrowsing}.
Together, these studies show that exposing meaningful structure can reduce reliance on element-by-element traversal and support more efficient non-visual navigation.

Structured navigation has also been explored directly for mobile screen readers~\cite{UbiComp2022SemanticNav,Zhang2021ScreenRecognition,Zhang2024DontConfuse}.
Most closely related to our work, ``Don't Confuse!'' groups nearby or visually similar GUI components into regions and uses these regions to reorganize the TalkBack navigation order~\cite{Zhang2024DontConfuse}.
Although this makes related elements more likely to be encountered consecutively, users still navigate the individual elements rather than the regions themselves.
A user-facing hierarchy requires a further step: functional groups must become navigable units, allowing users to move among groups and selectively enter one to inspect its elements.
This requires recovering functional structures such as cards, lists, and menus from general mobile GUIs.

\subsection{Recovering Structure in Mobile GUIs}

Recovering structure from general mobile GUIs has been studied as a problem of identifying relationships among otherwise separate UI elements.
Earlier approaches infer such structure from visual and spatial regularities.
Xie et al. use Gestalt cues such as proximity, similarity, connectivity, and continuity to recover perceptual groups including cards, menus, and lists~\cite{Xie2022PerceptualGrouping}, while Screen Parsing detects UI elements and predicts their relationships as a hierarchical UI model~\cite{Wu2021ScreenParsing}.
Related work models screen- and component-level semantics, element relationships, or semantic groups from visual and multimodal features~\cite{Fu2021PW2SS,Xiao2024SemanticGrouping,Li2021Screen2Vec,Wang2021Screen2Words}.
Together, these methods show that meaningful structure can be inferred even when it is not explicitly available in accessibility metadata.

Recent multimodal large language models (MLLMs) extend this capability by reasoning jointly over visual, textual, and spatial information in GUIs~\cite{You2024FerretUI,Wu2024MobileVLM,Wang2025MPGUI,Fan2024TreeOfLens}.
MP-GUI explicitly models spatial relationships such as containment and shared-parent relations to improve GUI understanding~\cite{Wang2025MPGUI}.
More directly, Tree-of-Lens constructs a hierarchical layout tree from GUI regions and uses it to interpret the content and spatial context around a user-indicated location~\cite{Fan2024TreeOfLens}.
These approaches suggest that modern models can recover not only individual elements but also relationships needed to reason about GUI structure.

However, recovered structure has largely served as an intermediate representation for GUI understanding, grounding, description, or downstream automation.
Even when hierarchy is screen reading, as in Tree-of-Lens, it supports the interpretation of a selected region rather than exposing the hierarchy itself for navigation.
The remaining challenge is therefore to recover function-aware groups at a granularity that can directly serve as user-facing navigation units, while preserving access to their underlying elements.

\section{Understanding Screen Reading and Design Needs}
\label{sec:screen-reading-design-needs}

Mobile screen readers, such as TalkBack~\cite{url:android-talkback} and VoiceOver~\cite{url:apple-voiceover}, enable BLV users to access mobile UIs through speech output, gesture commands, and sequential focus navigation.
In this section, we investigate current screen-reading experiences from two complementary perspectives.
We first conduct a quantitative audit of TalkBack to characterize recurring limitations of element-level reading.
We then conduct formative interviews with BLV users to understand how these limitations manifest in practice and what forms of support users need.
Finally, based on findings from both studies, we derive a set of design goals for hierarchical screen reading.

\subsection{Quantitative Audit}
\label{ssec:quantitative-audit}

We examine how TalkBack exposes and organizes UI elements during sequential navigation, and the resulting challenges for accessing functional content on mobile UIs.

\subsubsection{Corpus and Setup}

We constructed a corpus of 50 pages from 10 public Android apps spanning five categories: shopping, travel, social, finance, and entertainment.
Because our formative study focuses on BLV users in China, we selected two widely used apps in the Chinese mobile app market from each category.

For each app, we selected five key pages that represent its major functionality, including the homepage and pages directly reachable through homepage. 
This resulted in 50 pages covering a range of common mobile UI structures and interaction scenarios.
For each page, we recorded the screenshot, accessibility tree, and the focus trace produced by forward TalkBack navigation.

We further selected one representative interaction target on each page to measure navigation effort.
Targets were visible and actionable elements associated with the page's primary functionality.
We excluded system and back controls, persistent global navigation, advertisements, and elements whose activation involved payment, private information, or irreversible changes.
For consistency, the same apps, pages, and interaction targets are reused in the technical evaluation in Section~\ref{sec:technical-evaluation} to enable direct comparison under identical interface conditions.

\subsubsection{Measures}
\paragraph{Reference annotations}
Two authors, each with over three years of experience in mobile GUI research, independently annotated all pages for functional-unit membership and visible semantic actions.
A functional unit was defined as a group of spatially and semantically related elements that jointly represent a coherent piece of content or functionality, such as a product card or navigation bar.
For functional-unit annotation, we measured inter-annotator agreement using pairwise co-membership F1 $F1 = 0.92$, treating an element pair as positive if both elements were assigned to at least one common functional unit.
For semantic-action annotation, we report Cohen's kappa~\cite{Cohen1960Kappa} $\kappa = 0.81$.
Disagreements were resolved through discussion before computing the following measures.

\paragraph{Metrics}
We quantified four aspects of TalkBack's element-level reading:
\begin{itemize}
    \item \textit{Functional-Unit Interruptions} measure how often sequential navigation leaves a functional unit before all of its elements have been read and later returns to it. 
    For example, $A,A,B,A,B$ contains two interruptions, whereas $A,A,A,B,B$ contains none.
    \item \textit{Non-informative Focus Stops} count focus stops whose announcements convey no meaningful page-related content, function, state, or actionable purpose.
    These include empty or generic announcements (\eg \texttt{icon} without an identifying label), implementation-facing strings, isolated fragments, uninterpretable fragments (\eg \texttt{O1CN01pv1no7qe2iL1chua}), and duplicate announcements.
    Non-informative focus stops were independently labeled by two annotators. Disagreements were adjudicated by a third annotator through discussion until consensus was reached.
    \item \textit{Missed Semantic Actions} measure visible, user-meaningful actions in the reference annotation that are not reachable through TalkBack navigation. 
    We report both the number and proportion of annotated actions with no corresponding accessible focus stop.
    \item \textit{Target Navigation Distance} measure the number of forward-swipe gestures required to move TalkBack focus from the page's initial focus position to the predefined interaction target.
\end{itemize}

\subsubsection{Results.}
Table~\ref{tab:motivation-results} summarizes the audit results.
Functional-unit interruptions occurred on 17 of 50 pages (34.0\%), indicating that elements belonging to the same functional unit were not always encountered consecutively during TalkBack navigation.
Across all 1,858 focus stops, 535 (28.8\%) were judged non-informative.
TalkBack also failed to expose 161 annotated semantic actions (12.2\%).
Finally, reaching the predefined interaction targets required an average of 18.7 forward swipes.

These issues appeared across all five app categories, although their UI styles and designs varied by category.
Together, the results show that element-level screen reading can fragment functional structure, introduce substantial low-information content, omit meaningful actions, and require lengthy sequential navigation.
We next examine how BLV users experience these limitations in practice and what forms of structural support they need.

\begin{table*}[t]
\caption{Quantitative characterization of TalkBack reading across 50 pages and 50 interaction targets. Pages with interruptions, non-informative focus stops, and missed semantic actions are reported as count (percentage). Interruptions per page and target navigation distance are reported as mean $\pm$ SD; the latter is measured in forward-swipe gestures.}
\label{tab:motivation-results}
\centering
\small
\setlength{\tabcolsep}{4pt}
\renewcommand{\arraystretch}{1.15}
\begin{tabular*}{\textwidth}{@{\hspace{6pt}\extracolsep{\fill}}lccccc@{}}
\toprule
\shortstack{\textbf{App} \\ \textbf{Category}} &
\shortstack{\textbf{Pages with} \\ \textbf{Interruptions}} &
\shortstack{\textbf{Interruptions} \\ \textbf{per Page}} &
\shortstack{\textbf{Non-informative} \\ \textbf{Focus Stops}} &
\shortstack{\textbf{Missed Semantic} \\ \textbf{Actions}} &
\shortstack{\textbf{Target Navigation} \\ \textbf{Distance}} \\
\midrule
Shopping
& 3 (30.0\%)
& 0.4 $\pm$ 0.7
& 93 (23.4\%)
& 25 (8.0\%)
& 23.1 $\pm$ 16.2 \\
Travel
& 4 (40.0\%)
& 1.3 $\pm$ 1.8
& 199 (40.0\%)
& 51 (16.1\%)
& 29.7 $\pm$ 24.0 \\
Social
& 4 (40.0\%)
& 0.5 $\pm$ 0.7
& 112 (34.3\%)
& 18 (7.1\%)
& 12.2 $\pm$ 6.4 \\
Finance
& 5 (50.0\%)
& 0.8 $\pm$ 1.2
& 84 (23.3\%)
& 30 (13.6\%)
& 15.5 $\pm$ 17.4 \\
Entertainment
& 1 (10.0\%)
& 0.6 $\pm$ 1.9
& 47 (17.1\%)
& 37 (16.8\%)
& 12.8 $\pm$ 6.8 \\
\midrule
Overall
& 17 (34.0\%)
& 0.7 $\pm$ 1.4
& 535 (28.8\%)
& 161 (12.2\%)
& 18.7 $\pm$ 16.5 \\
\bottomrule
\end{tabular*}
\end{table*}

\subsection{Formative Interviews}
\label{ssec:formative-interviews}

We conducted formative interviews to understand how BLV users navigate and make sense of complex mobile interfaces with screen readers.

\subsubsection{Method.}
\label{ssec:formative-method}

\paragraph{Participants.}
We recruited ten BLV mobile screen-reader users (P1--P10).
Participants were 23--56 years old ($M=39.8$, $SD=10.4$), including four female and six male participants.
Eight participants were blind, and two reported light perception.
All participants regularly used screen readers on mobile devices.
Their everyday screen readers included VoiceOver\cite{url:apple-voiceover}, TalkBack\cite{url:android-talkback}, Dianming~\cite{url:dianming-android}, Baoyi~\cite{url:baoyi-yueting}, and Tiantan~\cite{url:tiantan-screenreader}, with screen-reader experience ranging from 5 to 17 years.
Table~\ref{table:3.2_participants} summarizes their demographics, visual status, mobile platforms, and screen-reader experience.

Participants were recruited through snowball sampling and in-person recruitment.
They received RMB~100 (approximately US\$15) per hour.
Before participation, we explained the study procedures, data recording, and voluntary nature of participation, and obtained informed consent from all participants.

\begin{table}[t]
    \centering
    \caption{Participant demographics and mobile screen-reader experience.}
    \label{table:3.2_participants}
    \small
    \renewcommand{\arraystretch}{1.12}

    \begin{tabular*}{\textwidth}{
        @{\hspace{6pt}\extracolsep{\fill}}
        c
        c
        c
        >{\centering\arraybackslash}p{2.8cm}
        c
        >{\centering\arraybackslash}p{2.8cm}
        c
        @{}
    }
        \toprule
        \textbf{ID}
        & \textbf{Age}
        & \textbf{Gender}
        & \textbf{Vision}
        & \textbf{Platform}
        & \textbf{Primary SR(s)}
        & \textbf{SR Experience} \\
        \midrule

        P1  & 51   & Female & Blind & iOS & VoiceOver & $\sim$14 years \\
        P2  & 27   & Female & Light perception & iOS     & VoiceOver  & 12 years         \\
        P3  & 56 & Female & Blind (late onset)                   & iOS     & VoiceOver         & 5 years          \\
        P4  & 39 & Male & Light perception                     & iOS     & VoiceOver         & 13 years         \\
        P5  & 35 & Male & Blind                                & Android & Baoyi, Dianming   & 17 years         \\
        P6  & 38 & Male & Blind                                & Android & TalkBack          & 14 years         \\
        P7  & 36 & Male & Blind                                & Android & Dianming, Tiantan & 14 years         \\
        P8  & 47 & Male & Blind                                & iOS     & VoiceOver         & $\sim$15 years   \\
        P9  & 46 & Female & Blind                                & iOS     & VoiceOver         & $\sim$16 years   \\
        P10 & 23 & Male & Blind                                & Android & Dianming          & 10 years         \\

        \bottomrule
    \end{tabular*}
\end{table}

\paragraph{Procedure.}
The semi-structured interviews proceeded in three stages.
We first asked participants about their everyday mobile use, screen-reader experience, frequently used apps, and strategies for exploring unfamiliar interfaces.
We then asked them to recall or demonstrate interactions they found particularly easy or difficult.
During these examples, we asked participants how they detected missing or incorrect information, when they repeated or restarted traversal, and when they abandoned a task or sought assistance.
Finally, we introduced the concept of hierarchical reading, in which users first access a functional unit and then decide whether to inspect its elements.
Participants discussed where such access might be useful, what information and navigation it should provide, and concerns about efficiency, errors, trust, and control.

\paragraph{Analysis.}
\label{ssec:formative-analysis}
The interviews were audio-recorded and transcribed alongside the researchers' field notes.
Two researchers independently coded the interview records, focusing on participants' screen-reading difficulties, coping strategies, and expectations for hierarchical access while allowing additional themes to emerge from the data.
They compared and consolidated their initial codes into a shared codebook, which was iteratively refined as the remaining interviews were analyzed.
Related codes were then grouped into higher-level themes, with disagreements resolved through discussion.

\subsubsection{Findings}
\label{ssec:formative-findings}

\paragraph{Breakdowns in Sequential Access.}
Participants described accessibility failures that extended beyond missing labels.
Object identity was frequently unclear: P2--P5 and P7--P10 described controls or content announced with missing or overly generic names.
Focus granularity was another recurring problem.
P2--P5, P7, and P8 encountered interfaces where information belonging to one object was split across multiple focus targets, or where several controls were merged into a single target.
Relationships could also disappear even when the underlying text was read correctly.
P2, P3, P7, P8, and P10 described cases where attributes could not be reliably associated with their objects, such as prices separated from product names or dates announced without their surrounding context.
P7, for example, preferred a product name and price to be presented together rather than as dispersed focus targets, while P10 described hearing ``the 19th'' without knowing which month it referred to.
Participants also reported controls that could be read but not activated (P4, P5, P7, P8, P10), as well as unreliable focus, selection, or state feedback (P2, P4, P5, P7, P8, P10).
Delayed feedback was another recurring problem, making it difficult to determine whether an action had taken effect, particularly in time-sensitive situations (P2, P4--P6, P8, P9).
Together, these accounts show that element-level accessibility could break at multiple points: object identity, grouping, operability, and feedback, even when individual content remained readable.

\paragraph{Workarounds and Hidden Effort.}
Participants had developed extensive strategies for working around accessibility breakdowns.
P1, P2, P7, P9, and P10 relied on remembered locations and familiar navigation paths to bypass repeated element-by-element exploration.
When native screen-reader output was insufficient, P2--P7 used built-in or third-party recognition features and sometimes combined them with repeated touching, swiping, or reopening an interface.
P2, for example, described turning on VoiceOver's Screen Recognition~\cite{url:apple-screen-recognition} and repeatedly exploring a page when part of the interface was not read correctly.
Others reduced navigation through search, sorting, or known shortcuts (P4--P7, P9, P10), switched between devices or screen readers (P4--P6, P8), or sought help from family members, volunteers, customer service, or other BLV users.
These strategies often allowed participants to complete otherwise difficult tasks, but shifted the cost of inaccessible structure to additional interaction, memory, tool switching, and external assistance.

\paragraph{Expectations for Hierarchical Access.}
Participants generally viewed hierarchical access as useful when it organized related information into functional units without removing access to details.
P2--P5 and P7--P10 wanted each functional unit to first identify what it represented, such as a product, store, or section, before presenting detailed attributes.
They also wanted information belonging to the same functional unit to be announced together rather than split across separate announcements.
P2, P4, P5, and P7--P10 favored a progressive pattern in which users could first hear a concise summary of a functional unit and then enter it when more detail was needed.
However, participants did not equate less information with better access.
P1, P4, P5, and P7--P10 stressed that decision-relevant details, such as price, material, specification, or state, should remain available.
They preferred shallow hierarchies with clear labels, particularly on unfamiliar or complex pages, while familiar interfaces should preserve learned navigation paths.
P2, P6--P8, and P10 further emphasized that users should decide which units to enter and which controls to activate rather than having the system make these choices automatically.
Across these responses, participants expected hierarchical access to provide an optional layer of organization while preserving complete information, familiar interaction, and direct user control.

\subsection{Design Considerations}
\label{ssec:design-considerations}

Findings from the audit and interviews lead to three design considerations (DCs) for a structure-aware screen reader.

\begin{description}[style=nextline]
    \item[DC1: Present the UI as Functional Units]
    When exploring a page or searching for a target, BLV users often do not need to hear every individual element.
    Reading functional units and their purposes can help users identify relevant content without listening to details unrelated to their goal.
    Screen readers should therefore present the UI as functional units in most scenarios to reduce unnecessary listening and support more efficient navigation.

    \item[DC2: Preserve User Control over Atomic Elements]
    Unit-level reading should not replace users' access to atomic elements.
    Users must remain able to inspect the original content and operate individual controls with familiar gestures.
    Therefore, hierarchical exploration should be designed to preserve this control by letting users decide when to expand a unit and which elements to inspect or activate.

    \item[DC3: Maintain Stable and Responsive Navigation]
    For BLV users, stable navigation supports reuse of learned interaction paths, while timely feedback helps them follow the effects of their actions.
    Inconsistent accessible structures can require renewed exploration, and delays can leave users uncertain about whether an action has taken effect.
    Navigation should therefore preserve familiar patterns and minimize waiting for feedback.
\end{description}

Together, these considerations call for unit-level understanding, control over atomic elements, and predictable, responsive access to both.
The next section describes how Spatial Screen Reader constructs and exposes this representation.

\section{Approach}
\label{sec:approach}

\subsection{Overview}

\begin{figure*}[t]
    \centering
    \includegraphics[width=\textwidth]{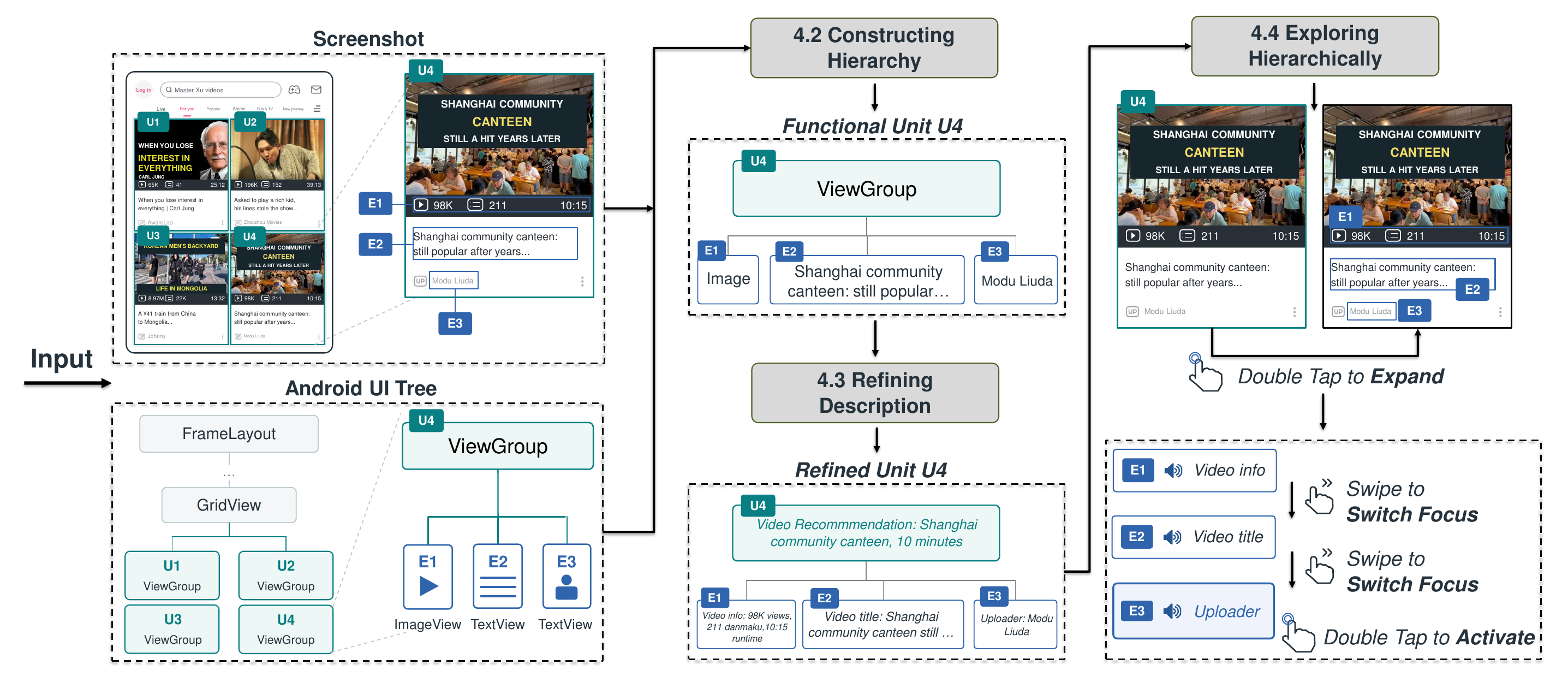}
    \caption{Overview of Spatial Screen Reader: (a) capture accessibility nodes, (b) build a function-aware hierarchy, (c) optionally refine labels with an LLM while preserving the hierarchy, and (d) explore functional units and activate their actions.}
    \label{fig:approach-overview}
    \Description{Four panels illustrate the workflow on a video recommendation interface. Accessibility nodes are grouped into a functional unit M1 containing actions A1 for opening the video, A2 for the creator, and A3 for more options. An optional LLM refines the unit label without changing the hierarchy. The user focuses on M1, expands its actions, and double-taps A3 to activate more options.}
\end{figure*}

\MIG takes Android accessibility metadata and spatial bounds as input to represent a mobile UI as a hierarchy of functional units, interactive elements, and readable content.
As shown in Figure~\ref{fig:approach-overview}, \MIG retrieves the page's accessibility nodes, including text, content descriptions, interaction properties, and coordinates.
It selects interactive targets and readable content, then identifies functional units using deterministic spatial and structural rules.
To this end, each node retains bounds for exploration and a spoken label; interactive nodes also link to corresponding actions in the underlying app.

Then, these nodes are organized into a hierarchical structure according to their containment relationships, with the page as the root.
Functional units contain related elements or smaller units, ordered from left to right within rows and from top to bottom across rows.
For interaction, \MIG adopts TalkBack's touch exploration and swipe navigation, allowing users to focus on nodes by touch or move between them with left and right swipes.
Double-tapping an expandable node replaces it with its immediate children for further exploration.
Original click actions are attached to interactive leaf nodes, which cannot be expanded, and users execute these actions with a double tap.
Users can thus inspect selected units in detail and activate individual controls to perform corresponding actions in the app.

\subsection{Hierarchy Construction}
\subsubsection{Preparing Atomic Elements}

Android provides an accessibility tree that reflects the current UI state~\cite{url:android-accessibility-node}, and \MIG serializes the captured tree as a static XML snapshot for hierarchy construction.
Each XML node element corresponds to one captured accessibility node, with its attributes recording text, content descriptions, interaction properties, and spatial bounds.
The nesting of these elements preserves the tree's parent-child relationships.

Both representations, however, contain redundant nodes and nodes that provide neither readable content nor an available action.
As reported in Table~\ref{tab:motivation-results}, 535 of 1,858 TalkBack focus stops (28.8\%) were non-informative.
\MIG therefore filters the captured nodes before using them to construct the hierarchy.

Specifically, \MIG filters nodes by their spatial bounds, interaction properties, and textual content.
As for the spatial bounds, it excludes nodes with no on-screen area.
For interaction properties, it retains all of the interactive nodes, excluding scrollable containers.
For textual content, it retains visible non-interactive nodes with text or content descriptions as content candidates.
Moreover, for interaction targets with identical bounds, \MIG keeps one representative, giving preference to nodes with readable labels or content descriptions.
Each selected element retains its spatial bounds and associated text for subsequent grouping and description generation.

\subsubsection{Identifying Functional Units}

After filtering non-informative nodes, \MIG groups the remaining atomic elements into functional units.
Inspired by the spatial grouping ability of sighted users~\cite{Xie2022PerceptualGrouping}, \MIG first identifies functional units through spatial containment and XML hierarchy.
For containment, when a larger node encloses smaller ones, the larger node inherently becomes a candidate functional unit containing those smaller nodes.
Besides, the XML structure provides additional grouping evidence: when two or more elements share a parent, that parent becomes a candidate unit.
Newly formed units can be grouped with their siblings in the same way, allowing functional units to form larger ones.

\begin{figure*}[t]
    \centering
    \begin{subfigure}[t]{0.48\textwidth}
        \centering
        \includegraphics[width=\linewidth]{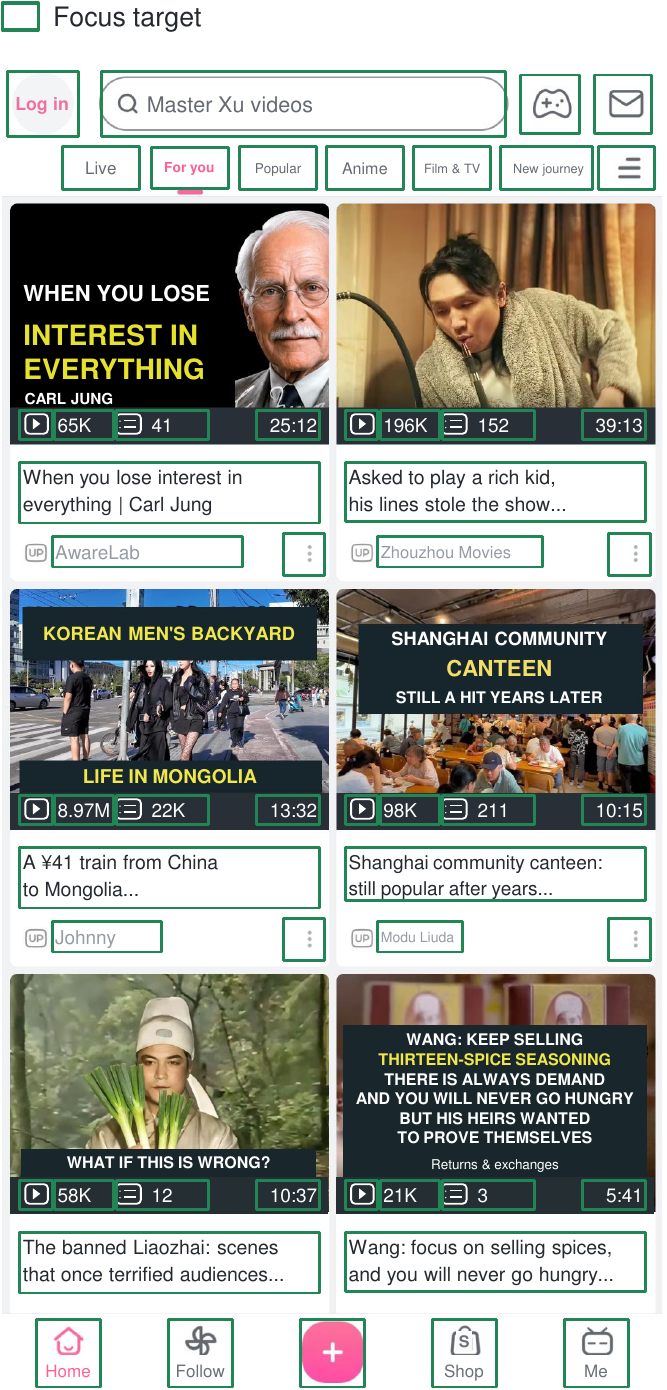}
        \caption{TalkBack}
        \label{fig:tree-compare-talkback}
    \end{subfigure}\hfill
    \begin{subfigure}[t]{0.48\textwidth}
        \centering
        \includegraphics[width=\linewidth]{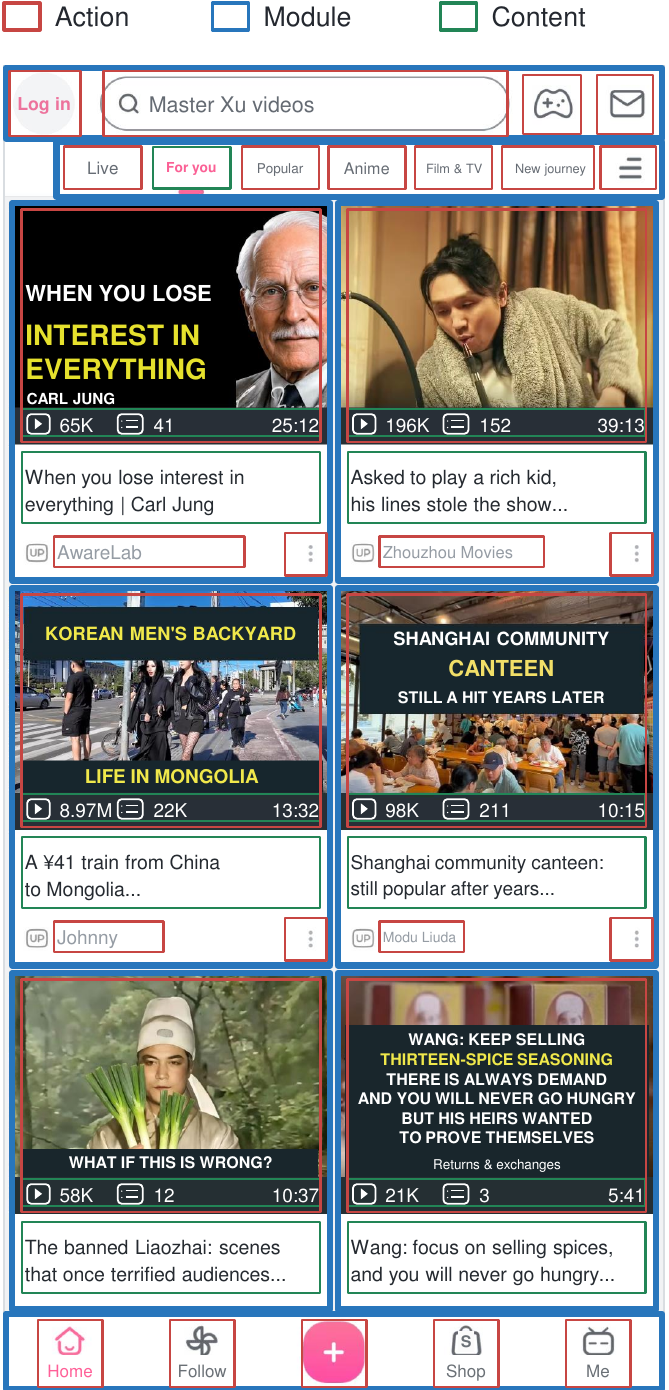}
        \caption{\MIG}
        \label{fig:tree-compare-seer}
    \end{subfigure}
    \caption{Comparison of the interface representations exposed by TalkBack and \MIG. TalkBack presents individual focus targets, whereas \MIG organizes actions and content into function-aware modules.}
    \Description{Two side-by-side panels show the same mobile video-feed page. The TalkBack panel outlines individual focus targets in green. The SeeR panel outlines actions in red, modules in blue, and content in green, showing how related elements are grouped into functional units.}
    \label{fig:tree-compare}
\end{figure*}

When spatial containment and XML hierarchy leave related elements ungrouped, \MIG uses horizontal and vertical alignment to form additional units.
This rule applies when adjacent elements or units form a row or column without an enclosing functional unit.
To avoid grouping unrelated elements, elements in a row must have similar heights, aligned top edges, and small horizontal gaps; elements in a column must have similar widths, aligned left edges, and small vertical gaps.
Finally, as shown in Figure~\ref{fig:tree-compare}, \MIG aggregates fragmented UI elements into units with explicit functions.

Feedback from our user study (Section~\ref{sec:user-study}) suggests that organizing related content into functional units helps users understand list content.

After identifying functional units, \MIG organizes the identified units and their constituent elements into a tree rooted at the current page.
Each unit becomes the parent of the elements it groups, while nested units are attached to their enclosing units according to the detected spatial or XML relationships.
Elements and units without a selected parent remain directly under the page root.
For each unit, its children are ordered from left to right within rows and from top to bottom across rows, with each unit's children kept consecutive during navigation.

Although the constructed hierarchy reflects the UI's functional structure, it may sometimes be too deep for efficient navigation.
This concern echoes participants' preference for shallow hierarchies in our formative study (Section~\ref{ssec:formative-findings}).
To help users avoid traversing unnecessary levels or hearing overly general information, \MIG simplifies the hierarchy through three rules.
Specifically, oversized containers, defined as those covering at least 80\% of the screen, are removed if they lack their own text or content descriptions.
Their children are attached directly to the next retained node.
Then, \MIG groups sibling elements or units under their shared XML parent for at most two rounds to avoid excessive depth.
Finally, units with fewer than three children are collapsed to expose their children's content or actions directly, reducing the number of levels users need to traverse.

\subsection{Refining Readable Description}

In the constructed hierarchy, both atomic elements and functional units may carry lengthy textual descriptions that make it difficult for BLV users to quickly grasp their content or purpose.
Therefore, \MIG prompts an LLM to compress descriptions from the available text and accessibility metadata into concise expressions that users can readily understand.
Given nodes at different levels encompass different amounts of contextual information and therefore require different degrees of compression.
Accordingly, as shown in Figure~\ref{fig:3_prompt}, \MIG applies three compression strategies for pages, functional units, and atomic elements.

At the page level, the root node of the hierarchy should provide an announcement that tells users which page they have entered and what content or functions they can explore.
Therefore, \MIG prompts an LLM to generate a page description based on the descriptions of the root node's direct children.
Specifically, the output should identify the page type and briefly summarize up to three of its main functions or content areas.
Then, whenever a page finishes loading, \MIG will automatically announce this description.

As for functional units, their descriptions should prioritize the core information users need to decide whether to expand a unit or move to the next one.
Although they are not atomic nodes, functional units are the most frequently visited nodes during hierarchical exploration, so their descriptions should be particularly concise.
Therefore, \MIG prompts an LLM to generate a concise name with one or two key attributes selected according to the application's domain.
For example, in e-commerce, a product summary retains its name, price, and sales volume so users can assess the product before opening its details.

Atomic elements typically suffer from missing or uninformative labels~\cite{Chen2020LabelDroid}.
For example, an icon may be labelled only \textit{button}, which does not convey its intended action.
Therefore, \MIG prompts an LLM to first identify whether an element has a concise, informative label.
Then, clear labels are retained, while missing or uninformative labels are regenerated using information from the element's parent and sibling nodes.
For example, the icon labelled only \textit{button} can be described as \textit{Open shopping cart} when the surrounding content indicates that action.

\begin{figure*}[t]
    \begin{tcbraster}[
        raster columns=3,
        raster equal height=rows,
        raster column skip=2.5mm
    ]
        \begin{promptbox}{1. Page Overview}
            Given the current page context and its major functional units,
            generate a concise and specific page name for the screen reader to
            announce when the user enters the page.

            \begin{itemize}[leftmargin=*,nosep]
                \item Infer the page function primarily from the page structure,
                functional units, and visible content, rather than relying only
                on the app or window name.
                \item Use the app name as contextual information, but do not
                return it alone unless the available evidence is insufficient.
                \item Prefer a specific description such as ``Shopping Home
                Page'' or ``Restaurant Details Page'' over a generic app name.
                \item Keep the page name short, stable, and easy to understand
                when spoken by a screen reader.
                \item When the available evidence is insufficient, use a
                conservative description rather than guessing.
                \item Return only the generated page name.
            \end{itemize}
        \end{promptbox}        \begin{promptbox}{2. Functional Unit Labeling}
            For each functional unit, generate a concise label that describes
            both its function and, when useful, its specific content.

            \begin{itemize}[leftmargin=*,nosep]
                \item Infer the unit's function from its child elements, their
                spatial relationships, and their collective purpose.
                \item Summarize the unit rather than concatenating the labels of
                its individual elements.
                \item Use consistent functional descriptions for structurally
                similar units while preserving differences in their content.
                \item Prefer labels such as ``Search Area,'' ``Filter Bar,''
                ``Product Recommendation: Milk,'' or ``Bottom Navigation.''
                \item Avoid inferring specific functions or categories that are
                not clearly supported by the available content.
            \end{itemize}
        \end{promptbox}        \begin{promptbox}{3. Atomic Action Label Refinement}
            Preserve an atomic element's original accessibility label when it is
            already clear and informative; refine it only when the label is
            missing, ambiguous, fragmented, or implementation-oriented.

            \begin{itemize}[leftmargin=*,nosep]
                \item Retain clear existing labels whenever possible.
                \item Rewrite concatenated, incomplete, or
                implementation-facing labels into natural descriptions suitable
                for speech.
                \item For unlabeled icons or images, infer their function
                conservatively from surrounding content, position, and available
                metadata.
                \item Preserve the key information in text-rich elements while
                removing redundant or low-level details.
                \item Do not expose internal identifiers, implementation names,
                or unsupported interpretations to users.
            \end{itemize}
        \end{promptbox}
    \end{tcbraster}
    \caption{Prompts for generating readable descriptions at the page,
    functional-unit, and atomic-action levels.}
    \Description{Three boxes arranged from left to right contain the prompts for
    generating a page overview, labeling functional units, and refining atomic
    action labels.}
    \label{fig:3_prompt}
\end{figure*}

\subsection{Supporting Hierarchical Exploration}

Hierarchical exploration should allow users to decide which details to inspect and which actions to perform.
Since the root node must be expanded to access the page's content, \MIG automatically announces its description and expands it to expose its direct children when the page is ready.
Deeper levels remain collapsed until users choose to expand the corresponding units.

To preserve familiar exploration patterns, \MIG retains TalkBack's touch exploration and left and right swipe gestures~\cite{url:talkback-gestures}, allowing users to focus on exposed nodes and hear their descriptions.
When user wants to expand a functional unit, double-tapping it will replace it with its direct children in the exposed sequence and automatically moves focus to the first child.
Since functional units and their constituent elements should retain their original actions, so the corresponding click actions are associated with interactive leaf nodes, which can be execute by double-tapping.

To further reduce learning effort, speech feedback should make the current focus, available operations, and the system's response clear to users.
Accordingly, \MIG announces the focused node's description and whether it can be expanded or activated.
For example, when a functional unit is focused, \MIG announces its description and indicates that it can be expanded. When user double-taps on it, \MIG announces ``expanded'' followed by the description of the first child to distinguish it from clicking an interactive leaf node.

\subsection{Stability and Responsiveness Enhancement}

Following DC3 in Section~\ref{ssec:design-considerations}, repeated visits to the same page should expose a consistent hierarchy so that users can reuse learned exploration paths.
Therefore, \MIG uses deterministic rules to keep grouping and navigation order as consistent as possible for the same captured page state.
Besides, since LLM-generated descriptions inevitably vary across requests, we lower the model's sampling temperature~\cite{url:deepseek-chat-api} and tailor prompts to different levels of the hierarchy to reduce this variation.

Given that dynamic changes are common in mobile UIs, the hierarchy should reflect changes in page content and structure while keeping interaction predictable.
Accordingly, \MIG continuously monitors the UI, recapturing the page and rebuilding the hierarchy after scrolling or a significant structural change.
During rebuilding, it temporarily blocks navigation and activation and provides a spoken loading message to prevent users from acting on an outdated hierarchy.
Once the updated hierarchy is ready, \MIG announces the page and resumes exploration.

As for responsiveness, LLM-based description generation inevitably introduces additional waiting time that should be minimized during page loading.
To reduce this delay, \MIG uses a small model (deepseek-v4-flash~\cite{url:deepseek-v4-flash}) for text compression and label generation, reducing the inference time of each request.
Moreover, instead of issuing separate requests for individual nodes, \MIG sends the hierarchy of the entire page in a single request and prompts the model to return the refined descriptions in a structured JSON format.

\section{Technical Evaluation}
\label{sec:technical-evaluation}
In this section, we evaluate three technical aspects of \MIG corresponding to its main design components:
1) whether it constructs the function-aware hierarchy accurately, completely, and efficiently;
2) whether the functional units and atomic elements are labeled by LLMs at acceptable runtime and monetary cost;
and 3) whether hierarchical navigation reduces the effort required to reach interaction targets.
All evaluations use the same 50-page corpus and interaction targets introduced in Section~\ref{ssec:quantitative-audit}, together with the human annotations as ground truth. 

\subsection{Function-Aware Hierarchy Construction}
\subsubsection{Setup and Measures}
We evaluate the constructed hierarchy along three dimensions: whether the identified functional units agree with human annotations, whether all the atomic actions are preserved, and how efficiently the hierarchy can be constructed.

\paragraph{Functional-unit identification}
We compare the generated hierarchy against the adjudicated functional-unit annotations introduced in Section~\ref{ssec:quantitative-audit}.
For each pair of atomic UI elements, we examine whether they share at least one functional unit in the generated hierarchy and in the human annotation.
We report pairwise precision, recall, and F1, macro-averaged across pages.
Precision reflects unrelated elements that are incorrectly grouped together, while recall reflects related elements that remain separated.
F1 summarizes the overall identification quality.

\paragraph{Action preservation}
We use the adjudicated semantic-action annotations introduced in Section~\ref{ssec:quantitative-audit} to examine whether the constructed hierarchy preserves access to atomic actions.
A semantic action is considered missed if no corresponding reachable atomic element is retained in the constructed hierarchy.
We manually verify the mappings between annotated actions and hierarchy nodes and report the number and percentage of missed semantic actions.

\paragraph{Construction latency}
We measure the time from receiving the parsed UI hierarchy (\ie the XML representation~\cite{url:UI_layout}) to completing construction of the function-aware hierarchy.
Semantic labeling and subsequent interaction processing are excluded.
Each page is measured five times after one warm-up run.
We report the median and 95th-percentile (P95) latency.

\subsubsection{Results}
Table~\ref{tab:hierarchy-construction-results} summarizes the effectiveness and efficiency of function-aware hierarchy construction.
Overall, \MIG achieves a pairwise precision of 0.996, recall of 0.711, and F1 of 0.769.
Precision remains at or above 0.98 across all app categories, whereas recall ranges from 0.52 to 0.85.
This consistent gap indicates that \MIG is conservative in constructing functional units: it rarely groups unrelated elements together, but sometimes leaves related elements separated.
This behavior is consistent with our design intent, as under-grouping primarily increases navigation effort, whereas incorrect grouping may introduce misleading relationships and confuse users.

Identification quality varies across app categories.
Social and shopping apps achieve F1 scores above 0.80, while travel apps obtain the lowest F1 of 0.59.
Our inspection shows that more travel pages rely on WebView-based~\cite{url:webview} interfaces, whose XML representations often expose atomic elements with limited meaningful container structure.
Such sparse structural cues make it more difficult for \MIG to recover the functional relationships among these elements.

The constructed hierarchies retain 85.2\% of the annotated semantic actions overall, with 196 of 1,324 actions missed.
The missed-action rate varies considerably across categories, from 3.5\% for social apps to 31.5\% for travel apps.
Further inspection shows that many of these missed actions arise from the same limitation: some actions exposed through the accessibility tree are absent from the XML hierarchy used for construction.

Finally, hierarchy construction introduces little computational overhead.
The median latency across all pages is 5.8ms, with a P95 latency of 21.5ms.
Even for travel apps, the slowest category, the median and P95 latencies are 9.7ms and 45.9ms, respectively.
Such latency is unlikely to introduce noticeable delay during screen reading.

\begin{table*}[t]
    \caption{Effectiveness and efficiency of function-aware hierarchy
    construction by app category. Functional-unit identification precision,
    recall, and F1 are macro-averaged across pages. Missed semantic actions are
    reported for \MIG as count (percentage of eligible gold
    actions). Latency is in milliseconds. Arrows indicate the preferred
    direction.}
    \label{tab:hierarchy-construction-results}
    \centering
    \small
    \setlength{\tabcolsep}{2pt}
    \renewcommand{\arraystretch}{1.15}
    \begin{tabular*}{\textwidth}{@{\hspace{6pt}\extracolsep{\fill}}lcccccc@{}}
        \toprule
        \multicolumn{1}{c}{\multirow{2}{*}{\textbf{App category}}} &
        \multicolumn{3}{c}{\textbf{Unit Identification Accuracy $\uparrow$}} &
        \multirow{2}{*}{\textbf{Missed Semantic Actions $\downarrow$}} &
        \multicolumn{2}{c}{\textbf{Latency (ms) $\downarrow$}} \\
        \cmidrule(lr){2-4}\cmidrule(lr){6-7}
        & \textbf{Precision} & \textbf{Recall} & \textbf{F1} &
        & \textbf{Median} & \textbf{P95} \\
        \midrule
        Shopping & 1.00 & 0.81 & 0.87 & 33 (10.6\%) & 5.8 & 20.8 \\
        Travel & 1.00 & 0.52 & 0.59 & 100 (31.5\%) & 9.7 & 45.9 \\
        Social & 1.00 & 0.85 & 0.91 & 9 (3.5\%) & 5.9 & 12.9 \\
        Finance & 1.00 & 0.66 & 0.69 & 39 (17.6\%) & 5.8 & 18.0 \\
        Entertainment & 0.98 & 0.73 & 0.79 & 15 (6.8\%) & 5.4 & 10.1 \\
        \midrule
        Overall & 1.00 & 0.71 & 0.77 & 196 (14.8\%) & 5.8 & 21.5 \\
        \bottomrule
    \end{tabular*}
\end{table*}

\subsection{Readable Description Generation}
\subsubsection{Setup and Measures}
We evaluate whether LLM-based labeling improves the informativeness of screen-reader output and quantify the associated runtime and token overhead.
We use deepseek-v4-flash, the same model deployed in the subsequent user study, to ensure consistency between the technical and user evaluations.
We select this model because it is readily accessible to users in China and supports long-context processing, making it suitable for our target deployment environment.

\paragraph{Semantic quality}
We apply the same criterion and annotation protocol for non-informative focus stops as in Section~\ref{ssec:quantitative-audit}.
Two annotators independently judge each generated announcement as informative or non-informative without knowing its source condition, with disagreements resolved by a third annotator through discussion.
We report the number and percentage of non-informative focus stops and compare the overall rate with the TalkBack baseline reported in Table~\ref{tab:motivation-results}.

\paragraph{Generation overhead}
We measure the end-to-end latency of each LLM request used to generate readable semantics.
For each page, we also record the number of prompt and completion tokens and report their average usage.
Latency is summarized using the median and 95th percentile (P95) across pages.

\subsubsection{Results}
As shown in Table~\ref{tab:generating-semantics-results}, across all pages, 82 generated focus stops (5.1\%) are judged non-informative, compared with 28.8\% of TalkBack focus stops in the quantitative audit.
This corresponds to an 82.1\% relative reduction, showing that semantic generation substantially reduces low-information content in screen-reader output.
The better annotated elements include aggregated functional units, images, and icons whose meanings depend on contextual information.

However, requesting LLMs introduces substantial latency.
Across all pages, the median LLM latency is 3.4s and the P95 latency is 5.8s.
Each page uses an average of 2,535.4 prompt tokens and 273.8 completion tokens. 
Under the API pricing of the used model, this corresponds to less than CNY~0.01 (less than \$0.002) per page.
Thus, the primary cost of semantic generation is time rather than money: while the monetary cost is negligible, LLM inference remains the main source of runtime overhead.

\begin{table*}[t]
    \caption{Readable semantic generation quality and overhead by app
    category. Non-informative focus stops are reported as count (percentage).
    Latency measures end-to-end LLM request time in seconds; token usage is averaged per page.
    Lower values are better.}
    \label{tab:generating-semantics-results}
    \centering
    \small
    \setlength{\tabcolsep}{3pt}
    \renewcommand{\arraystretch}{1.15}
    \begin{tabular*}{\textwidth}{@{\hspace{6pt}\extracolsep{\fill}}lccccc@{}}
        \toprule
        \textbf{App category} &
        \makecell{\textbf{Non-informative} \\ \textbf{Focus Stops}} &
        \textbf{Median Latency (s)} &
        \textbf{P95 Latency (s)} &
        \textbf{Prompt Tokens} &
        \textbf{Completion Tokens} \\
        \midrule
        Shopping & 12 (3.2\%) & 4.0 & 5.6 & 2,680.2 & 360.8 \\
        Travel & 8 (2.7\%) & 3.3 & 8.3 & 2,656.8 & 252.8 \\
        Social & 18 (6.2\%) & 3.7 & 4.4 & 2,378.2 & 239.6 \\
        Finance & 37 (11.3\%) & 3.4 & 4.2 & 2,380.0 & 238.3 \\
        Entertainment & 7 (2.3\%) & 3.2 & 5.2 & 2,581.6 & 277.5 \\
        \midrule
        Overall & 82 (5.1\%) & 3.4 &
        5.8 & 2,535.4 & 273.8 \\
        \bottomrule
    \end{tabular*}
\end{table*}

\subsection{Hierarchical Navigation}
\subsubsection{Setup and Measures}
We evaluate whether the function-aware hierarchy reduces the navigation effort required to reach a target element.
We use the same 50 interaction targets introduced in Section~\ref{ssec:quantitative-audit}, with identical page states and starting positions for TalkBack and \MIG.

For each target, we measure the minimum number of sequential navigation operations required to move from the predefined starting position to the target.
In TalkBack, each forward swipe advances the focus to the next element.
In \MIG, navigation proceeds among sibling nodes at the current level, while entering a functional unit requires an additional expansion operation.
All navigation paths are manually verified to ensure that each operation is valid and that the final node corresponds to the intended interaction target.
We report the mean and standard deviation across targets, together with the relative reduction of \MIG over TalkBack.

\subsubsection{Results}
Table~\ref{tab:hierarchical-navigation-results} summarizes the target navigation distance. 
Across all 50 targets, \MIG requires an average of 8.8 navigation steps ($SD=6.2$), compared with 18.7 ($SD=16.5$) for TalkBack, corresponding to a 52.7\% reduction. 
The mean paired reduction is 9.8 steps (95\% bootstrap CI~\cite{Efron1979Bootstrap}: [4.9, 15.0]).
\MIG requires fewer navigation steps for 38 of the 50 targets, the same number for 4 targets, and more steps for 8 targets.
A two-sided Wilcoxon signed-rank test~\cite{Wilcoxon1945} further confirms a significant reduction in navigation distance ($W=108.5$, $p<.001$).

The reduction is consistent across all five app categories, ranging from 25.8\% for finance apps to 74.4\% for travel apps.
These reductions are particularly pronounced when targets appear deep in TalkBack's flat reading sequence, as the hierarchy allows users to first navigate to the relevant functional unit before inspecting its elements.
Overall, these results show that hierarchical navigation substantially shortens the sequential access path to interaction targets.

\begin{table}[t]
    \caption{Target navigation distance by app category. Values are mean
    $\pm$ SD. The percentage following each \MIG value indicates its relative reduction from TalkBack.
    Lower values are better.}
    \label{tab:hierarchical-navigation-results}
    \centering
    \small
    \setlength{\tabcolsep}{10pt}
    \renewcommand{\arraystretch}{1.25}
    \begin{tabular*}{0.9\linewidth}{@{\hspace{6pt}\extracolsep{\fill}}lcc@{}}
        \toprule
        \multicolumn{1}{c}{\textbf{App category}} & \textbf{TalkBack} & \textbf{\MIG} \\
        \midrule
        Shopping & 23.1 $\pm$ 16.2 & 11.7 $\pm$ 4.9 (49.4\% $\downarrow$) \\
        Travel & 29.7 $\pm$ 24.0 & 7.6 $\pm$ 4.7 (74.4\% $\downarrow$) \\
        Social & 12.2 $\pm$ 6.4 & 7.3 $\pm$ 4.9 (40.2\% $\downarrow$) \\
        Finance & 15.5 $\pm$ 17.4 & 11.5 $\pm$ 9.7 (25.8\% $\downarrow$) \\
        Entertainment & 12.8 $\pm$ 6.8 & 6.0 $\pm$ 3.4 (53.1\% $\downarrow$) \\
        \midrule
        Overall & 18.7 $\pm$ 16.5 & 8.8 $\pm$ 6.2 (52.7\% $\downarrow$) \\
        \bottomrule
    \end{tabular*}
\end{table}

\section{User Study}
\label{sec:user-study}

We conducted a within-subjects study with ten BLV screen-reader users to compare \MIG with conventional linear reading.
The study examined whether function-aware hierarchical reading helped users understand UI functions and complete tasks.
It also assessed the effort required to learn and use hierarchical reading, together with participants' experiences and preferences.

\subsection{Participants}
\label{ssec:user-study-participants}

The same ten BLV participants (P1--P10) who completed the formative interviews in Section~\ref{ssec:formative-interviews} took part in the user study. 
Participants were 23--56 years old ($M=39.8$, $SD=10.4$), including four female and six male participants; eight were blind and two reported light perception. 
All were regular mobile screen-reader users, with 5--17 years of experience, and their everyday screen readers included VoiceOver\cite{url:apple-voiceover}, TalkBack\cite{url:android-talkback}, Baoyi\cite{url:baoyi-yueting}, Dianming\cite{url:dianming-android}, and Tiantan\cite{url:tiantan-screenreader}. 
Table~\ref{table:3.2_participants} summarizes their demographics, visual status, mobile platforms, and screen-reader experience. 
The formative interviews were completed before participants had any hands-on experience with \MIG, so their formative responses were not influenced by prototype use. 
Participants received RMB~100 (approximately US\$15) per hour. 
Interviews were held at locations chosen by the participants, and both interviews and experimental sessions were conducted in quiet environments.
All participants provided written informed consent, and their privacy was protected throughout data collection and analysis.

\subsection{Procedure}
\label{ssec:user-study-procedure}

\subsubsection{Study Design and Tasks.}
We conducted a within-subjects comparison between \MIG and TalkBack using four single-page tasks in Meituan~\cite{url:meituan} and JD~\cite{url:app:JD}, two widely used Chinese mobile commerce apps.
The tasks represented two common forms of mobile interaction: comparing information across repeated items and locating a specific functional entry.
Specifically, participants were asked to select the nearest cinema, enter the beverage-topping section, select the cheapest product, and open the after-sales order entry.
Each task began on a predefined page where the information or control required to complete the task was already present, so participants did not need to navigate from another page to reach the target.
The same task instruction and page state were used in both conditions.
Participants attempted each administered task with both \MIG and TalkBack.
The beverage-topping task was not administered to P1, P3, or P9, yielding 37 paired task attempts across the ten participants (Table~\ref{tab:user-task-completion}).
To mitigate condition-order effects, five participants used \MIG first and the other five used TalkBack first.

\subsubsection{Apparatus and Practice.}
To provide a consistent experimental environment, all participants used the same Android phone (Xiaomi 13, Android 16) with both \MIG and TalkBack installed, regardless of the screen reader or mobile platform they used in everyday life.
TalkBack was used with its default navigation settings, and the speech rate was kept consistent across the two conditions.

Before the formal experiment, the interviewer introduced and demonstrated the navigation and activation operations for both conditions.
For \MIG, this included moving among functional units, expanding a unit, navigating its underlying elements, and activating a control.
The tutorial used a separate Dianping restaurant page, where participants practiced locating the merchant rating; this page and task were not included in the four formal tasks.
All participants, including those who did not routinely use TalkBack, then practiced both reading modes freely until they reported that they were ready to proceed.
The tutorial and practice lasted approximately 7.5 minutes in total.

\subsubsection{Interaction Tasks and Measures.}
At the beginning of each task, the interviewer stated the target and opened the corresponding predefined page, and informed the participant that the task could be completed within the current page.
Participants then explored the page using the assigned reading mode and attempted to reach the specified target.
A task was considered completed when the participant reached the predefined goal; otherwise, an administered task was recorded as attempted but not completed.

With participants' consent, we recorded the experimental sessions and logged task completion, task time, and the number of navigation nodes visited.
Task time was measured from the start to the end of each task attempt.
For paired tasks attempted in both conditions, regardless of completion, we calculated each participant's mean number of nodes visited and task time, along with paired changes (\MIG minus TalkBack), and then averaged these measures across participants with at least one paired task.
For inferential comparison, we used two-sided Wilcoxon signed-rank tests to compare participant-level mean node counts and task times between \MIG and TalkBack, with Holm correction for the two comparisons.

\subsubsection{Questionnaire and Interview.}
After completing the interaction tasks, participants rated eight comparative statements on a five-point Likert scale from 1 (\textit{strongly disagree}) to 5 (\textit{strongly agree}).
For reporting, we reverse-scored Q2, Q4, Q6, and Q8 as $6-x$ and presented all eight items with positive wording, so higher scores consistently indicate more favorable evaluations of \MIG.
Figure~\ref{fig:user-questionnaire} summarizes participants' responses on region understanding, ease of using controls, action confidence, ease of learning, operation convenience, perceived information completeness, preference for hierarchical reading on unfamiliar or complex pages, and its perceived advantage on simple or familiar pages.
Participants also indicated whether they generally preferred TalkBack, \MIG, or had no clear preference.

In the subsequent semi-structured interview, participants explained their ratings, compared the two reading modes across tasks, described difficulties encountered during the experiment, and discussed what they found most helpful and what they would most like to improve.
Participants who remained interested could additionally try multi-page tasks after the formal study; these exploratory interactions were used only as qualitative material and were excluded from the quantitative comparison.

\subsection{Results and Analysis}
\label{ssec:user-study-results}

\subsubsection{Task Performance}

\paragraph{Task Completion.}
Table~\ref{tab:user-task-completion} summarizes task completion for each participant, task, and reading condition.
Across all administered trials, participants completed 36/37 tasks (97.3\%) with \MIG and 34/37 tasks (91.9\%) with TalkBack.
Among the 37 tasks administered in both conditions, 33 were completed with both systems, three only with \MIG, and one only with TalkBack; none remained incomplete in both conditions.
Thus, \MIG achieved a similar overall task-completion rate while changing how participants accessed the page.
Completion nevertheless varied across tasks, indicating that hierarchical access was not uniformly advantageous for every type of target.
\begin{table*}[t]
    \caption{Task completion by participant, task, and condition.
    \ding{51} indicates completion; \ding{55} indicates that the task was attempted but not completed; $\bigcirc$ indicates that the task was not executed.
    TB = TalkBack.}
    \label{tab:user-task-completion}
    \centering
    \small
    \setlength{\tabcolsep}{3pt}
    \renewcommand{\arraystretch}{1.15}
    \begin{tabular*}{\textwidth}{@{\hspace{6pt}}p{3.2em}*{8}{>{\centering\arraybackslash}p{\dimexpr(\textwidth-6pt-3.2em-16\tabcolsep)/8\relax}}@{}}
        \toprule
        \multirow{2}{*}[-0.6ex]{\textbf{ID}} &
        \multicolumn{2}{c}{\textbf{Nearest cinema}} &
        \multicolumn{2}{c}{\textbf{Beverage toppings}} &
        \multicolumn{2}{c}{\textbf{Cheapest item}} &
        \multicolumn{2}{c}{\textbf{After-sales order}} \\
        \cmidrule(lr){2-3}\cmidrule(lr){4-5}\cmidrule(lr){6-7}\cmidrule(lr){8-9}
        & \textbf{TB} & \textbf{\MIG} & \textbf{TB} & \textbf{\MIG} &
        \textbf{TB} & \textbf{\MIG} & \textbf{TB} & \textbf{\MIG} \\
        \midrule
                P1 & \ding{55} & \ding{51} & $\bigcirc$ & $\bigcirc$ & \ding{55} & \ding{51} & \ding{51} & \ding{55} \\
        P2 & \ding{51} & \ding{51} & \ding{51} & \ding{51} & \ding{51} & \ding{51} & \ding{51} & \ding{51} \\
        P3 & \ding{55} & \ding{51} & $\bigcirc$ & $\bigcirc$ & \ding{51} & \ding{51} & \ding{51} & \ding{51} \\
        P4 & \ding{51} & \ding{51} & \ding{51} & \ding{51} & \ding{51} & \ding{51} & \ding{51} & \ding{51} \\
        P5 & \ding{51} & \ding{51} & \ding{51} & \ding{51} & \ding{51} & \ding{51} & \ding{51} & \ding{51} \\
        P6 & \ding{51} & \ding{51} & \ding{51} & \ding{51} & \ding{51} & \ding{51} & \ding{51} & \ding{51} \\
        P7 & \ding{51} & \ding{51} & \ding{51} & \ding{51} & \ding{51} & \ding{51} & \ding{51} & \ding{51} \\
        P8 & \ding{51} & \ding{51} & \ding{51} & \ding{51} & \ding{51} & \ding{51} & \ding{51} & \ding{51} \\
        P9 & \ding{51} & \ding{51} & $\bigcirc$ & $\bigcirc$ & \ding{51} & \ding{51} & \ding{51} & \ding{51} \\
        P10 & \ding{51} & \ding{51} & \ding{51} & \ding{51} & \ding{51} & \ding{51} & \ding{51} & \ding{51} \\
        \midrule
        \textbf{Overall} &
        \textbf{8/10} & \textbf{10/10} &
        \textbf{7/7} & \textbf{7/7} &
        \textbf{9/10} & \textbf{10/10} &
        \textbf{10/10} & \textbf{9/10} \\
        \bottomrule
    \end{tabular*}
\end{table*}

\paragraph{Interaction Efficiency.}
Table~\ref{tab:user-interaction-cost} reports participant-level mean node counts and task times for paired tasks attempted in both conditions, regardless of completion.
Participants averaged 9.7 nodes per task with \MIG ($SD=1.4$), compared with 17.9 nodes ($SD=2.8$) with TalkBack, corresponding to a mean participant-level reduction of 8.2 nodes (45.8\%).
This reduction was significant according to a two-sided Wilcoxon signed-rank test ($W=0$, Holm-adjusted $p=.004$).
Mean task time was 38.3 seconds ($SD=12.9$) with \MIG and 58.2 seconds ($SD=27.3$) with TalkBack, with an average paired reduction of 19.9 seconds (34.2\%).
Task time was also significantly lower with \MIG ($W=4$, Holm-adjusted $p=.014$).
All ten participants with paired observations visited fewer nodes with \MIG, while eight of ten also spent less time on average.
Although the magnitude of the differences varied across participants, the overall pattern indicates that hierarchical access reduced navigation and generally shortened task attempts while maintaining a similar overall completion rate.
\begin{table*}[t]
    \caption{Interaction performance by participant and condition.
    Order summarizes the system sequence for each participant (S = \MIG; T = TalkBack).
    Nodes and time are means across paired tasks attempted in both conditions, regardless of completion.
    Their Overall values report the mean (SD) across participant means.
    Paired changes are \MIG minus TalkBack; the Overall change averages participant-level paired differences.}
    \label{tab:user-interaction-cost}
    \centering
    \small
    \setlength{\tabcolsep}{2pt}
    \renewcommand{\arraystretch}{1.15}
    \begin{tabular*}{\textwidth}{@{\hspace{6pt}\extracolsep{\fill}}l*{7}{c}@{}}
        \toprule
        \multirow{2}{*}[-0.6ex]{\textbf{ID}} & \multirow{2}{*}[-0.6ex]{\textbf{Order}} & \multicolumn{2}{c}{\textbf{TalkBack}} &
        \multicolumn{2}{c}{\textbf{\MIG}} &
        \multicolumn{2}{c}{\textbf{Change}} \\
        \cmidrule(lr){3-4}\cmidrule(lr){5-6}\cmidrule(lr){7-8}
        & & \textbf{Nodes} &
        \textbf{Time (s)} & \textbf{Nodes} &
        \textbf{Time (s)} & \textbf{$\Delta$ nodes} &
        \textbf{$\Delta$ time (s)} \\
        \midrule
                                P1 & S$\rightarrow$T & 19.7 & 108.3 & 9.3 & 54.3 & $-10.3$ & $-54.0$ \\
        P2 & S$\rightarrow$T & 17.8 & 74.8 & 11.0 & 32.5 & $-6.8$ & $-42.3$ \\
        P3 & S$\rightarrow$T & 14.7 & 81.3 & 11.0 & 56.0 & $-3.7$ & $-25.3$ \\
        P4 & T$\rightarrow$S & 16.8 & 37.0 & 9.3 & 44.3 & $-7.5$ & $+7.3$ \\
        P5 & T$\rightarrow$S & 19.8 & 46.8 & 10.0 & 32.3 & $-9.8$ & $-14.5$ \\
        P6 & T$\rightarrow$S & 13.8 & 34.0 & 7.3 & 24.0 & $-6.5$ & $-10.0$ \\
        P7 & S$\rightarrow$T & 23.8 & 38.0 & 11.5 & 27.5 & $-12.3$ & $-10.5$ \\
        P8 & S$\rightarrow$T & 17.5 & 61.3 & 10.0 & 55.3 & $-7.5$ & $-6.0$ \\
        P9 & T$\rightarrow$S & 17.3 & 79.0 & 7.7 & 30.7 & $-9.7$ & $-48.3$ \\
        P10 & T$\rightarrow$S & 17.8 & 21.5 & 10.0 & 25.8 & $-7.8$ & $+4.3$ \\
        \midrule
        \textbf{Overall} & --- & \textbf{17.9 (2.8)} & \textbf{58.2 (27.3)} & \textbf{9.7 (1.4)} & \textbf{38.3 (12.9)} & \textbf{-8.2} & \textbf{-19.9} \\
        \bottomrule
    \end{tabular*}
\end{table*}

\subsubsection{Navigation Behavior}
\label{ssec:user-navigation-behavior}
\paragraph{Functional grouping reduced the immediate search space.}
By aggregating related elements into functional units, \MIG reduced the number of candidates participants needed to consider at each stage of navigation.
Participants could first move among functional units and enter only those that appeared relevant, rather than inspect every underlying element individually.
Participants repeatedly described this organization as reducing unnecessary scanning.
P2 explained that when element-by-element exploration became overwhelming, the hierarchy helped by telling her what a region was for before she inspected its contents.
P9 described a similar process through an analogy to organizing objects into categories: ``Leaving everything on the surface makes it harder to search,'' whereas opening a larger category first made it easier to find smaller items within it.
Together with the consistently lower node counts observed in the task trials, these accounts show that functional grouping reduced sequential search by making irrelevant regions skippable as units rather than requiring participants to inspect their elements individually.

\paragraph{Hierarchy reduced breadth through selective summaries.}
\MIG reduced navigation breadth not only by grouping related elements into fewer functional units, but also by presenting only a concise summary of each unit at the current level.
These summaries retained a small set of representative attributes, such as product prices and cinema distances, while deferring additional details to deeper levels.
This organization was particularly useful when the exposed information aligned with what participants needed for the task.
P2 identified the product task as the one she liked most because \MIG ``told me the main information and helped me focus on the key points.''
P9 similarly described the product name and price as the key information and found the complete element-level presentation ``too verbose.''
Importantly, deferred information remained accessible when needed.
As P3 explained, ``I can open an item and listen when I need the details.''
Thus, the hierarchy reduced breadth by keeping the current level concise while preserving access to detailed information at deeper levels.

\paragraph{Hierarchy depth added navigation and decision costs.}
The information deferred from the current level remained accessible, but reaching it required additional navigation.
When a target was nested inside a functional unit, participants first had to decide which unit to enter and then perform an expansion action.
Clear descriptions reduced this decision cost by signaling what was available inside the unit.
Ambiguous descriptions made deeper navigation more difficult because participants could not determine whether entering a unit would bring them closer to the target.
For example, in the after-sales task, participants were asked to locate the ``after-sales order'' entry on the JD account page.
\MIG first exposed a broader order-related functional unit, whereas TalkBack directly announced ``after-sales'' as participants moved through the controls.
P3 found TalkBack more convenient in this case because the target could be recognized immediately from its label.
Across the tasks, hierarchical navigation therefore involved a breadth--depth trade-off.
Grouping and summarization reduced the amount of information exposed at the current level, while deeper access introduced additional interaction and decision costs.

\subsubsection{User Experience and Feedback}

\begin{figure*}[t]
    \centering
    \includegraphics[width=\linewidth]{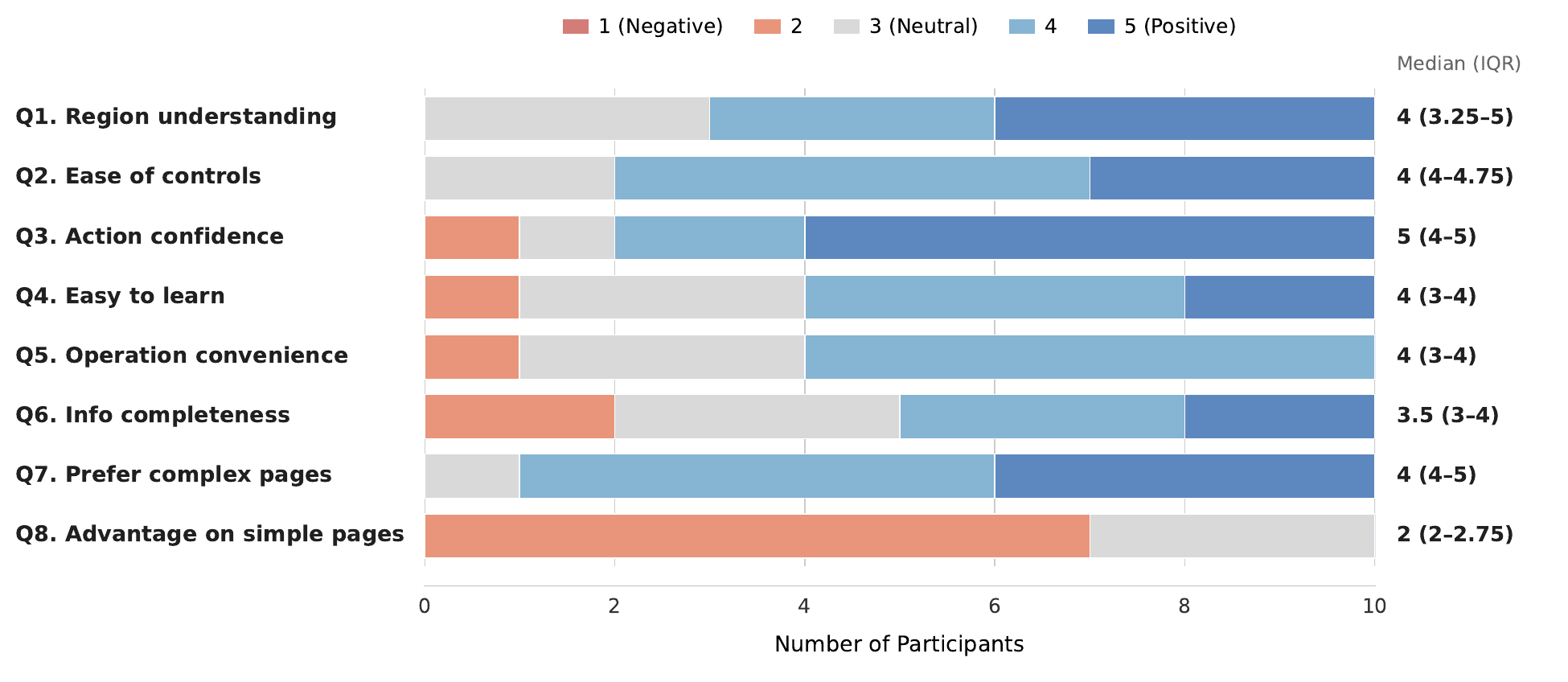}
    \caption{Participants' responses to the post-task five-point Likert-scale questionnaire.
    Higher scores indicate more favorable evaluations of \MIG; values on the right show the median and IQR.}
    \Description{Eight horizontal stacked bars show the distribution of ten participants' ratings from one to five. Median ratings are four for region understanding, ease of controls, ease of learning, operation convenience, and preference on complex pages; five for action confidence; 3.5 for information completeness; and two for advantage on simple pages.}
    \label{fig:user-questionnaire}
\end{figure*}

Figure~\ref{fig:user-questionnaire} summarizes participants' ratings of hierarchical reading.
Ratings were generally favorable for region understanding (Q1, median = 4), ease of using controls (Q2, median = 4), action confidence (Q3, median = 5), ease of learning (Q4, median = 4), and operation convenience (Q5, median = 4).

\paragraph{Participants valued structured and concise access.}
Participants often associated \MIG with a clearer organization of page content and less unnecessary listening.
Rather than traversing all elements individually at the same level, they could first understand the purpose of a functional unit and inspect its contents when necessary.
This organization also gave some participants greater confidence that they were acting within the intended part of the interface.
Overall, their ratings and comments suggest that the perceived benefit came from making information easier to organize and act on, rather than simply shortening the spoken output.

\paragraph{Conciseness required recoverable detail.}
Participants also recognized that summarization could hide information they might later need.
P6 noted that on first use, it was difficult to know whether important information had been omitted, and that this trust would require longer-term use to establish.
Others were more accepting when deferred details remained easy to recover.
For example, P7 considered a short product summary sufficient for initial comparison because the full details could still be inspected before purchase.
Thus, concise access was most acceptable when information was deferred rather than removed.

\paragraph{Experiences depended on context and user control.}
Participants generally favored hierarchical reading for unfamiliar or complex pages (Q7, median = 4), while ratings indicated limited perceived advantage on simple or familiar pages (Q8, median = 2).
Several participants relied on familiar locations and learned navigation paths, reducing the value of an additional hierarchy in routine interactions.
P2 therefore suggested allowing users to choose between more structured guidance and a faster familiar mode depending on their experience.
These responses indicate that participants viewed hierarchical reading as a complementary access mode whose value depended on the page, the task, and the user's familiarity.

\section{Discussion}
\label{sec:discussion}

The key observation from our studies is that being able to access a control does not necessarily mean understanding its function.
When a reading sequence obscures which elements belong together, users can hardly understand their function before reconstructing their relationships through repeated exploration.
In this work, \MIG makes these relationships explicit through function-aware grouping and unit summaries, but navigating this structure also introduces extra effort, particularly when users already know their target.
This raises a design question: how should functional units be summarized, and when and how should these summaries be presented to support understanding while minimizing unnecessary interaction effort?

\subsection{Navigation-Oriented Unit Summaries}
\label{ssec:unit-summary-design}

A functional unit can be grouped correctly yet summarized in a way that offers little guidance about whether to explore it.
Participants valued hearing a unit's purpose before inspecting its controls and using that overview to skip unrelated content (Section~\ref{ssec:user-navigation-behavior}).
\textit{A unit summary should communicate functional relevance, rather than merely shorten the element sequence.}
It should explain what the unit offers and what its controls act on, helping users decide whether to enter it.

Summaries should also distinguish units that share a function.
For example, describing every product card as ``product information and purchase options'' conveys their common purpose but offers little basis for choosing among them.
We therefore suggest retaining an identifying name along with a few function-relevant details.
In this example, ``a butterfly watch for \$10.01'' would identify the item and provide a price for comparison.
The key criterion for a summary is thus whether it supports a meaningful navigation choice.
We acknowledge that the details best suited to supporting such choices remain to be determined, as our study did not compare alternative summary strategies.

\subsection{Adaptive Hierarchical and Direct Access}
\label{ssec:hierarchical-direct-access}

Even an informative summary adds effort when users do not need the information it provides.
P4 cautioned against excessive hierarchy, and P5 described the effort of moving between levels.
On a familiar interface, users may already remember the precise location to find their targets, making additional hierarchical navigation unnecessary.
Designers should therefore allow users to skip summaries and access elements directly while retaining unit-level guidance on demand.

In fact, familiarity does not necessarily imply a complete understanding of an interface.
In our study, participants sometimes had difficulty locating similar targets even on interfaces they were familiar with.
This observation suggests that users should be able to switch between hierarchical and direct access at any point during exploration, with their current focus preserved.
This would allow them to request more or less guidance without restarting exploration.

\subsection{Stable and Responsive Structural Navigation}
\label{ssec:stable-responsive-navigation}

As reflected in DC3 (Section~\ref{ssec:design-considerations}), many participants expected a stable structure and consistent descriptions.
This is because, once a BLV user learns how to interact with a UI, remembering familiar paths can make subsequent interactions much more faster.
A stable accessibility structure therefore helps BLV users draw on prior experience without reconstructing their understanding each time.
However, frequent UI revisions~\cite{DBLP:conf/icst/LiCWH0WL17} and variable LLM-generated descriptions can disrupt this familiarity.
Such variability can hardly be avoid by the post-hoc proceedure of screen readers.

Therefore, \textit{only app designers can preserve unaffected functional units and descriptions.}
Specifically, we strongly recommend that designers, who know their apps best, update the accessibility structure and descriptions upon completing the visual UI and keep them current through subsequent iterations.
We hope that the output of \MIG can serve as a useful reference for developers during this process.
In addition, we hope that support from the community and platforms will help maintain this continuity.
Keeping this structure accurate and stable would help minimize repeated exploration and support consistent mobile use.

\subsection{Limitations and Threats to Validity}
\label{ssec:limitations}

First, the navigation strategy limits how our metrics should be interpreted.
Sequential TalkBack traversal supports exploration of unfamiliar screens but covers only one form of screen-reader interaction.
Experienced BLV users may instead use touch exploration, control-type shortcuts, search, or custom gestures.
Other access methods, such as voice control and magnification, involve different interaction costs.
Thus, Target Navigation Distance describes an idealized sequential path, not the full range of actual user behavior.

The study design introduces further validity concerns.
Reusing the formative-study sample may affect internal validity.
The same ten participants took part in both studies, potentially increasing their familiarity with the design concept before the comparative evaluation.
Additionally, participants may have evaluated the hierarchical reader more favorably because the interviewers were also its developers.

Finally, our evaluations cover 50 fixed pages from 10 Android apps and ten participants performing four single-page tasks.
This scope limits generalization to other platforms, dynamic pages, and cross-app tasks.
Therefore, larger-scale studies are recommended to examine the performance of hierarchical screen readers across different scenarios and applications.

\section{Conclusion}
\label{sec:conclusion}
This paper presented Spatial Screen Reader (\MIG), which organizes mobile UI elements into a function-aware hierarchy and supports navigation through unit summaries and access to underlying elements.
Compared with existing sequential readers, \MIG can reduce sequential navigation steps by 52.7\%.
Participants' accounts further suggest that unit summaries facilitate understanding of UI functions, although navigating across hierarchy levels may require additional actions.

These findings position spatial organization as accessibility information that apps and platforms should explicitly expose and maintain.
Screen readers are highly recommended to make this structure available while allowing users to switch between hierarchical and direct access according to the task and their familiarity with the interface.
Supporting this choice would help users obtain an overview when needed and retain convenient access to individual controls and details.

\bibliographystyle{ACM-Reference-Format}
\bibliography{ui_testing}

\end{document}